
%
\catcode`@=11 
%
%
%

\font\fourteenrm=cmr10 scaled\magstep2
\font\twelverm=cmr10 scaled\magstep1
\font\ninerm=cmr9            \font\sixrm=cmr6

\font\fourteenbf=cmbx10 scaled\magstep2
\font\twelvebf=cmbx10 scaled\magstep1
\font\ninebf=cmbx9            \font\sixbf=cmbx6
\font\seventeeni=cmmi10 scaled\magstep3     \skewchar\seventeeni='177
\font\fourteeni=cmmi10 scaled\magstep2      \skewchar\fourteeni='177
\font\twelvei=cmmi10 scaled\magstep1        \skewchar\twelvei='177
\font\ninei=cmmi9                           \skewchar\ninei='177
\font\sixi=cmmi6                            \skewchar\sixi='177
\font\seventeensy=cmsy10 scaled\magstep3    \skewchar\seventeensy='60
\font\fourteensy=cmsy10 scaled\magstep2     \skewchar\fourteensy='60
\font\twelvesy=cmsy10 scaled\magstep1       \skewchar\twelvesy='60
\font\ninesy=cmsy9                          \skewchar\ninesy='60
\font\sixsy=cmsy6                           \skewchar\sixsy='60

\font\fourteenex=cmex10 scaled\magstep2
\font\twelveex=cmex10 scaled\magstep1

\font\fourteensl=cmsl10 scaled\magstep2
\font\twelvesl=cmsl10 scaled\magstep1
\font\ninesl=cmsl9

\font\fourteenit=cmti10 scaled\magstep2
\font\twelveit=cmti10 scaled\magstep1
\font\twelvett=cmtt10 scaled\magstep1
\font\twelvecp=cmcsc10 scaled\magstep1
\font\tencp=cmcsc10
\newfam\cpfam
%
%
\newcount\f@ntkey            \f@ntkey=0
\def\samef@nt{\relax \ifcase\f@ntkey \rm \or\oldstyle \or\or
         \or\it \or\sl \or\bf \or\tt \or\caps \fi }
\def\fourteenpoint{\relax
    \textfont0=\fourteenrm          \scriptfont0=\tenrm
    \scriptscriptfont0=\sevenrm
     \def\rm{\fam0 \fourteenrm \f@ntkey=0 }\relax
    \textfont1=\fourteeni           \scriptfont1=\teni
    \scriptscriptfont1=\seveni
     \def\oldstyle{\fam1 \fourteeni\f@ntkey=1 }\relax
    \textfont2=\fourteensy          \scriptfont2=\tensy
    \scriptscriptfont2=\sevensy
    \textfont3=\fourteenex     \scriptfont3=\fourteenex
    \scriptscriptfont3=\fourteenex
    \def\it{\fam\itfam \fourteenit\f@ntkey=4 }\textfont\itfam=\fourteenit
    \def\sl{\fam\slfam \fourteensl\f@ntkey=5 }\textfont\slfam=\fourteensl
    \scriptfont\slfam=\tensl
    \def\bf{\fam\bffam \fourteenbf\f@ntkey=6 }\textfont\bffam=\fourteenbf
    \scriptfont\bffam=\tenbf     \scriptscriptfont\bffam=\sevenbf
    \def\tt{\fam\ttfam \twelvett \f@ntkey=7 }\textfont\ttfam=\twelvett
    \h@big=11.9\p@{} \h@Big=16.1\p@{} \h@bigg=20.3\p@{} \h@Bigg=24.5\p@{}
    \def\caps{\fam\cpfam \twelvecp \f@ntkey=8 }\textfont\cpfam=\twelvecp
    \setbox\strutbox=\hbox{\vrule height 12pt depth 5pt width\z@}
    \samef@nt}
\def\twelvepoint{\relax
    \textfont0=\twelverm          \scriptfont0=\ninerm
    \scriptscriptfont0=\sixrm
     \def\rm{\fam0 \twelverm \f@ntkey=0 }\relax
    \textfont1=\twelvei           \scriptfont1=\ninei
    \scriptscriptfont1=\sixi
     \def\oldstyle{\fam1 \twelvei\f@ntkey=1 }\relax
    \textfont2=\twelvesy          \scriptfont2=\ninesy
    \scriptscriptfont2=\sixsy
    \textfont3=\twelveex          \scriptfont3=\twelveex
    \scriptscriptfont3=\twelveex
    \def\it{\fam\itfam \twelveit \f@ntkey=4 }\textfont\itfam=\twelveit
    \def\sl{\fam\slfam \twelvesl \f@ntkey=5 }\textfont\slfam=\twelvesl
    \scriptfont\slfam=\ninesl
    \def\bf{\fam\bffam \twelvebf \f@ntkey=6 }\textfont\bffam=\twelvebf
    \scriptfont\bffam=\ninebf     \scriptscriptfont\bffam=\sixbf
    \def\tt{\fam\ttfam \twelvett \f@ntkey=7 }\textfont\ttfam=\twelvett
    \h@big=10.2\p@{}
    \h@Big=13.8\p@{}
    \h@bigg=17.4\p@{}
    \h@Bigg=21.0\p@{}
    \def\caps{\fam\cpfam \twelvecp \f@ntkey=8 }\textfont\cpfam=\twelvecp
    \setbox\strutbox=\hbox{\vrule height 10pt depth 4pt width\z@}
    \samef@nt}
\def\tenpoint{\relax
    \textfont0=\tenrm          \scriptfont0=\sevenrm
    \scriptscriptfont0=\fiverm
    \def\rm{\fam0 \tenrm \f@ntkey=0 }\relax
    \textfont1=\teni           \scriptfont1=\seveni
    \scriptscriptfont1=\fivei
    \def\oldstyle{\fam1 \teni \f@ntkey=1 }\relax
    \textfont2=\tensy          \scriptfont2=\sevensy
    \scriptscriptfont2=\fivesy
    \textfont3=\tenex          \scriptfont3=\tenex
    \scriptscriptfont3=\tenex
    \def\it{\fam\itfam \tenit \f@ntkey=4 }\textfont\itfam=\tenit
    \def\sl{\fam\slfam \tensl \f@ntkey=5 }\textfont\slfam=\tensl
    \def\bf{\fam\bffam \tenbf \f@ntkey=6 }\textfont\bffam=\tenbf
    \scriptfont\bffam=\sevenbf     \scriptscriptfont\bffam=\fivebf
    \def\tt{\fam\ttfam \tentt \f@ntkey=7 }\textfont\ttfam=\tentt
    \def\caps{\fam\cpfam \tencp \f@ntkey=8 }\textfont\cpfam=\tencp
    \setbox\strutbox=\hbox{\vrule height 8.5pt depth 3.5pt width\z@}
    \samef@nt}
%
%
%
%
\newdimen\h@big  \h@big=8.5\p@
\newdimen\h@Big  \h@Big=11.5\p@
\newdimen\h@bigg  \h@bigg=14.5\p@
\newdimen\h@Bigg  \h@Bigg=17.5\p@
\def\big#1{{\hbox{$\left#1\vbox to\h@big{}\right.\n@space$}}}
\def\Big#1{{\hbox{$\left#1\vbox to\h@Big{}\right.\n@space$}}}
\def\bigg#1{{\hbox{$\left#1\vbox to\h@bigg{}\right.\n@space$}}}
\def\Bigg#1{{\hbox{$\left#1\vbox to\h@Bigg{}\right.\n@space$}}}
%
%
%
\normalbaselineskip = 20pt plus 0.2pt minus 0.1pt
\normallineskip = 1.5pt plus 0.1pt minus 0.1pt
\normallineskiplimit = 1.5pt
\newskip\normaldisplayskip
\normaldisplayskip = 20pt plus 5pt minus 10pt
\newskip\normaldispshortskip
\normaldispshortskip = 6pt plus 5pt
\newskip\normalparskip
\normalparskip = 6pt plus 2pt minus 1pt
\newskip\skipregister
\skipregister = 5pt plus 2pt minus 1.5pt
\newif\ifsingl@    \newif\ifdoubl@
\newif\iftwelv@    \twelv@true
\def\singlespace{\singl@true\doubl@false\spaces@t}
\def\doublespace{\singl@false\doubl@true\spaces@t}
\def\normalspace{\singl@false\doubl@false\spaces@t}
\def\Tenpoint{\tenpoint\twelv@false\spaces@t}
\def\Twelvepoint{\twelvepoint\twelv@true\spaces@t}
\def\spaces@t{\relax%
 \iftwelv@ \ifsingl@\subspaces@t3:4;\else\subspaces@t1:1;\fi%
 \else \ifsingl@\subspaces@t3:5;\else\subspaces@t4:5;\fi \fi%
 \ifdoubl@ \multiply\baselineskip by 5%
 \divide\baselineskip by 4 \fi \unskip}
\def\subspaces@t#1:#2;{
      \baselineskip = \normalbaselineskip
      \multiply\baselineskip by #1 \divide\baselineskip by #2
      \lineskip = \normallineskip
      \multiply\lineskip by #1 \divide\lineskip by #2
      \lineskiplimit = \normallineskiplimit
      \multiply\lineskiplimit by #1 \divide\lineskiplimit by #2
      \parskip = \normalparskip
      \multiply\parskip by #1 \divide\parskip by #2
      \abovedisplayskip = \normaldisplayskip
      \multiply\abovedisplayskip by #1 \divide\abovedisplayskip by #2
      \belowdisplayskip = \abovedisplayskip
      \abovedisplayshortskip = \normaldispshortskip
      \multiply\abovedisplayshortskip by #1
        \divide\abovedisplayshortskip by #2
      \belowdisplayshortskip = \abovedisplayshortskip
      \advance\belowdisplayshortskip by \belowdisplayskip
      \divide\belowdisplayshortskip by 2
      \smallskipamount = \skipregister
      \multiply\smallskipamount by #1 \divide\smallskipamount by #2
      \medskipamount = \smallskipamount \multiply\medskipamount by 2
      \bigskipamount = \smallskipamount \multiply\bigskipamount by 4 }
\def\normalbaselines{ \baselineskip=\normalbaselineskip
   \lineskip=\normallineskip \lineskiplimit=\normallineskip
   \iftwelv@\else \multiply\baselineskip by 4 \divide\baselineskip by 5
     \multiply\lineskiplimit by 4 \divide\lineskiplimit by 5
     \multiply\lineskip by 4 \divide\lineskip by 5 \fi }
\Twelvepoint  
\interlinepenalty=50
\interfootnotelinepenalty=5000
\predisplaypenalty=9000
\postdisplaypenalty=500
\hfuzz=1pt
\vfuzz=0.2pt
%
%
%
\def\pagecontents{
   \ifvoid\topins\else\unvbox\topins\vskip\skip\topins\fi
   \dimen@ = \dp255 \unvbox255
   \ifvoid\footins\else\vskip\skip\footins\footrule\unvbox\footins\fi
   \ifr@ggedbottom \kern-\dimen@ \vfil \fi }
\def\makeheadline{\vbox to 0pt{ \skip@=\topskip
      \advance\skip@ by -12pt \advance\skip@ by -2\normalbaselineskip
      \vskip\skip@ \line{\vbox to 12pt{}\the\headline} \vss
      }\nointerlineskip}
\def\makefootline{\baselineskip = 1.5\normalbaselineskip
                 \line{\the\footline}}
\newif\iffrontpage
\newif\ifletterstyle
\newif\ifp@genum
\def\nopagenumbers{\p@genumfalse}
\def\pagenumbers{\p@genumtrue}
\pagenumbers
\newtoks\paperheadline
\newtoks\letterheadline
\newtoks\letterfrontheadline
\newtoks\lettermainheadline
\newtoks\paperfootline
\newtoks\letterfootline
\newtoks\date
\footline={\ifletterstyle\the\letterfootline\else\the\paperfootline\fi}
\paperfootline={\hss\iffrontpage\else\ifp@genum\tenrm\folio\hss\fi\fi}
\letterfootline={\hfil}
\headline={\ifletterstyle\the\letterheadline\else\the\paperheadline\fi}
\paperheadline={\hfil}
\letterheadline{\iffrontpage\the\letterfrontheadline
     \else\the\lettermainheadline\fi}
\lettermainheadline={\rm\ifp@genum page \ \folio\fi\hfil\the\date}
\def\monthname{\relax\ifcase\month 0/\or January\or February\or
   March\or April\or May\or June\or July\or August\or September\or
   October\or November\or December\else\number\month/\fi}
\date={\monthname\ \number\day, \number\year}
\countdef\pagenumber=1  \pagenumber=1
\def\advancepageno{\global\advance\pageno by 1
   \ifnum\pagenumber<0 \global\advance\pagenumber by -1
    \else\global\advance\pagenumber by 1 \fi \global\frontpagefalse }
\def\folio{\ifnum\pagenumber<0 \romannumeral-\pagenumber
           \else \number\pagenumber \fi }
\def\footrule{\dimen@=\prevdepth\nointerlineskip
   \vbox to 0pt{\vskip -0.25\baselineskip \hrule width 0.35\hsize \vss}
   \prevdepth=\dimen@ }
\newtoks\foottokens
\foottokens={\Tenpoint\singlespace}
\newdimen\footindent
\footindent=24pt
\def\vfootnote#1{\insert\footins\bgroup  \the\foottokens
   \interlinepenalty=\interfootnotelinepenalty \floatingpenalty=20000
   \splittopskip=\ht\strutbox \boxmaxdepth=\dp\strutbox
   \leftskip=\footindent \rightskip=\z@skip
   \parindent=0.5\footindent \parfillskip=0pt plus 1fil
   \spaceskip=\z@skip \xspaceskip=\z@skip
   \Textindent{$ #1 $}\footstrut\futurelet\next\fo@t}
\def\Textindent#1{\noindent\llap{#1\enspace}\ignorespaces}
\def\footnote#1{\attach{#1}\vfootnote{#1}}

\def\foot{\attach\footsymbolgen\vfootnote{\footsymbol}}
\let\footsymbol=\star
\newcount\lastf@@t           \lastf@@t=-1
\newcount\footsymbolcount    \footsymbolcount=0
\newif\ifPhysRev
\def\footsymbolgen{\relax \ifPhysRev \iffrontpage \NPsymbolgen\else
      \PRsymbolgen\fi \else \NPsymbolgen\fi
   \global\lastf@@t=\pageno \footsymbol }
\def\NPsymbolgen{\ifnum\footsymbolcount<0 \global\footsymbolcount=0\fi
   {\iffrontpage \else \advance\lastf@@t by 1 \fi
    \ifnum\lastf@@t<\pageno \global\footsymbolcount=0
     \else \global\advance\footsymbolcount by 1 \fi }
   \ifcase\footsymbolcount \fd@f\star\or \fd@f\dagger\or \fd@f\ast\or
    \fd@f\ddagger\or \fd@f\natural\or \fd@f\diamond\or \fd@f\bullet\or
    \fd@f\nabla\else \fd@f\dagger\global\footsymbolcount=0 \fi }
\def\fd@f#1{\xdef\footsymbol{#1}}
\def\PRsymbolgen{\ifnum\footsymbolcount>0 \global\footsymbolcount=0\fi
      \global\advance\footsymbolcount by -1
      \xdef\footsymbol{\sharp\number-\footsymbolcount} }
\def\space@ver#1{\let\@sf=\empty \ifmmode #1\else \ifhmode
   \edef\@sf{\spacefactor=\the\spacefactor}\unskip${}#1$\relax\fi\fi}
\def\attach#1{\space@ver{\strut^{\mkern 2mu #1} }\@sf\ }
%
%
%
\newcount\chapternumber      \chapternumber=0
\newcount\sectionnumber      \sectionnumber=0
\newcount\equanumber         \equanumber=0
\let\chapterlabel=0
\newtoks\chapterstyle        \chapterstyle={\Number}
\newskip\chapterskip         \chapterskip=\bigskipamount
\newskip\sectionskip         \sectionskip=\medskipamount
\newskip\headskip            \headskip=8pt plus 3pt minus 3pt
\newdimen\chapterminspace    \chapterminspace=15pc
\newdimen\sectionminspace    \sectionminspace=10pc
\newdimen\referenceminspace  \referenceminspace=25pc
\def\chapterreset{\global\advance\chapternumber by 1
   \ifnum\the\equanumber<0 \else\global\equanumber=0\fi
   \sectionnumber=0 \makel@bel}
\def\makel@bel{\xdef\chapterlabel{%
\the\chapterstyle{\the\chapternumber}.}}
\def\sectionlabel{\number\sectionnumber \quad }
\def\alphabetic#1{\count255='140 \advance\count255 by #1\char\count255}
\def\Alphabetic#1{\count255='100 \advance\count255 by #1\char\count255}
\def\Roman#1{\uppercase\expandafter{\romannumeral #1}}
\def\roman#1{\romannumeral #1}
\def\Number#1{\number #1}
\def\unnumberedchapters{\let\makel@bel=\relax \let\chapterlabel=\relax
\let\sectionlabel=\relax \equanumber=-1 }
\def\titlestyle#1{\par\begingroup \interlinepenalty=9999
     \leftskip=0.02\hsize plus 0.23\hsize minus 0.02\hsize
     \rightskip=\leftskip \parfillskip=0pt
     \hyphenpenalty=9000 \exhyphenpenalty=9000
     \tolerance=9999 \pretolerance=9000
     \spaceskip=0.333em \xspaceskip=0.5em
     \iftwelv@\fourteenpoint\else\twelvepoint\fi
   \noindent #1\par\endgroup }
\def\spacecheck#1{\dimen@=\pagegoal\advance\dimen@ by -\pagetotal
   \ifdim\dimen@<#1 \ifdim\dimen@>0pt \vfil\break \fi\fi}
\def\chapter#1{\par \penalty-300 \vskip\chapterskip
   \spacecheck\chapterminspace
   \chapterreset \titlestyle{\chapterlabel \ #1}
   \nobreak\vskip\headskip \penalty 30000
   \wlog{\string\chapter\ \chapterlabel} }

\def\section#1{\par \ifnum\the\lastpenalty=30000\else
   \penalty-200\vskip\sectionskip \spacecheck\sectionminspace\fi
   \wlog{\string\section\ \chapterlabel \the\sectionnumber}
   \global\advance\sectionnumber by 1  \noindent
   {\caps\enspace\chapterlabel \sectionlabel #1}\par
   \nobreak\vskip\headskip \penalty 30000 }
\def\subsection#1{\par
   \ifnum\the\lastpenalty=30000\else \penalty-100\smallskip \fi
   \noindent\undertext{#1}\enspace \vadjust{\penalty5000}}

\def\undertext#1{\vtop{\hbox{#1}\kern 1pt \hrule}}
\def\APPENDIX#1#2{\par\penalty-300\vskip\chapterskip
   \spacecheck\chapterminspace \chapterreset \xdef\chapterlabel{#1}
   \titlestyle{APPENDIX #2} \nobreak\vskip\headskip \penalty 30000
   \wlog{\string\Appendix\ \chapterlabel} }
\def\Appendix#1{\APPENDIX{#1}{#1}}
\def\appendix{\APPENDIX{A}{}}
%
%
%
\def\eqname#1{\relax \ifnum\the\equanumber<0
     \xdef#1{{\rm(\number-\equanumber)}}\global\advance\equanumber by -1
    \else \global\advance\equanumber by 1
      \xdef#1{{\rm(\chapterlabel \number\equanumber)}} \fi}

\def\eqn#1{\eqno\eqname{#1}#1}

\def\eqinsert#1{\noalign{\dimen@=\prevdepth \nointerlineskip
   \setbox0=\hbox to\displaywidth{\hfil #1}
   \vbox to 0pt{\vss\hbox{$\!\box0\!$}\kern-0.5\baselineskip}
   \prevdepth=\dimen@}}
%

%

%

%
%
\def\GENITEM#1;#2{\par \hangafter=0 \hangindent=#1
    \Textindent{$ #2 $}\ignorespaces}
\outer\def\newitem#1=#2;{\gdef#1{\GENITEM #2;}}
\newdimen\itemsize                \itemsize=30pt
\newitem\item=1\itemsize;
\newitem\sitem=1.75\itemsize;     
\newitem\ssitem=2.5\itemsize;     
\outer\def\newlist#1=#2&#3&#4;{\toks0={#2}\toks1={#3}%
   \count255=\escapechar \escapechar=-1
   \alloc@0\list\countdef\insc@unt\listcount     \listcount=0
   \edef#1{\par
      \countdef\listcount=\the\allocationnumber
      \advance\listcount by 1
      \hangafter=0 \hangindent=#4
      \Textindent{\the\toks0{\listcount}\the\toks1}}
   \expandafter\expandafter\expandafter
    \edef\c@t#1{begin}{\par
      \countdef\listcount=\the\allocationnumber \listcount=1
      \hangafter=0 \hangindent=#4
      \Textindent{\the\toks0{\listcount}\the\toks1}}
   \expandafter\expandafter\expandafter
    \edef\c@t#1{con}{\par \hangafter=0 \hangindent=#4 \noindent}
   \escapechar=\count255}
\def\c@t#1#2{\csname\string#1#2\endcsname}
\newlist\point=\Number&.&1.0\itemsize;
\newlist\subpoint=(\alphabetic&)&1.75\itemsize;
\newlist\subsubpoint=(\roman&)&2.5\itemsize;
\newlist\cpoint=\Roman&.&1.0\itemsize;
%

%
%
%
\newcount\referencecount     \referencecount=0
\newif\ifreferenceopen       \newwrite\referencewrite
\newtoks\rw@toks
\def\NPrefmark#1{\attach{\scriptscriptstyle [ #1 ] }}
\let\PRrefmark=\attach
\def\CErefmark#1{\attach{\scriptstyle  #1 ) }}
\def\refmark#1{\relax\ifPhysRev\PRrefmark{#1}\else\NPrefmark{#1}\fi}
\def\crefmark#1{\relax\CErefmark{#1}}
\def\refend{\refmark{\number\referencecount}}
\newcount\lastrefsbegincount \lastrefsbegincount=0
\def\refsend{\refmark{\count255=\referencecount
   \advance\count255 by-\lastrefsbegincount
   \ifcase\count255 \number\referencecount
   \or \number\lastrefsbegincount,\number\referencecount
   \else \number\lastrefsbegincount-\number\referencecount \fi}}
\def\crefsend{\crefmark{\count255=\referencecount
   \advance\count255 by-\lastrefsbegincount
   \ifcase\count255 \number\referencecount
   \or \number\lastrefsbegincount,\number\referencecount
   \else \number\lastrefsbegincount-\number\referencecount \fi}}
\def\refch@ck{\chardef\rw@write=\referencewrite
   \ifreferenceopen \else \referenceopentrue
   \immediate\openout\referencewrite=referenc.texauxil \fi}
%
{\catcode`\^^M=\active 
  \gdef\obeyendofline{\catcode`\^^M\active \let^^M\ }}%
%
{\catcode`\^^M=\active 
  \gdef\ignoreendofline{\catcode`\^^M=5}}
{\obeyendofline\gdef\rw@start#1{\def\t@st{#1} \ifx\t@st\blankend%
\endgroup \@sf \relax \else \ifx\t@st\bl@nkend \endgroup \@sf \relax%
\else \rw@begin#1
\backtotext
\fi \fi } }
{\obeyendofline\gdef\rw@begin#1
{\def\n@xt{#1}\rw@toks={#1}\relax%
\rw@next}}
\def\blankend{}
{\obeylines\gdef\bl@nkend{
}}
\newif\iffirstrefline  \firstreflinetrue
\def\rwr@teswitch{\ifx\n@xt\blankend \let\n@xt=\rw@begin %
 \else\iffirstrefline \global\firstreflinefalse%
\immediate\write\rw@write{\noexpand\obeyendofline \the\rw@toks}%
\let\n@xt=\rw@begin%
      \else\ifx\n@xt\rw@@d \def\n@xt{\immediate\write\rw@write{%
        \noexpand\ignoreendofline}\endgroup \@sf}%
             \else \immediate\write\rw@write{\the\rw@toks}%
             \let\n@xt=\rw@begin\fi\fi \fi}
\def\rw@next{\rwr@teswitch\n@xt}
\def\rw@@d{\backtotext} \let\rw@end=\relax
\let\backtotext=\relax

\newdimen\refindent     \refindent=30pt
\def\refitem#1{\par \hangafter=0 \hangindent=\refindent \Textindent{#1}}
\def\REFNUM#1{\space@ver{}\refch@ck \firstreflinetrue%
 \global\advance\referencecount by 1 \xdef#1{\the\referencecount}}
\def\refnum#1{\space@ver{}\refch@ck \firstreflinetrue%
 \global\advance\referencecount by 1 \xdef#1{\the\referencecount}\refend}

\def\REF#1{\REFNUM#1%
 \immediate\write\referencewrite{%
 \noexpand\refitem{#1.}}%
\begingroup\obeyendofline\rw@start}
\def\ref{\refnum\?%
 \immediate\write\referencewrite{\noexpand\refitem{\?.}}%
\begingroup\obeyendofline\rw@start}
\def\Ref#1{\refnum#1%
 \immediate\write\referencewrite{\noexpand\refitem{#1.}}%
\begingroup\obeyendofline\rw@start}
\def\REFS#1{\REFNUM#1\global\lastrefsbegincount=\referencecount
\immediate\write\referencewrite{\noexpand\refitem{#1.}}%
\begingroup\obeyendofline\rw@start}
\newcount\figurecount     \figurecount=0
\newif\iffigureopen       \newwrite\figurewrite
\def\figch@ck{\chardef\rw@write=\figurewrite \iffigureopen\else
   \immediate\openout\figurewrite=figures.texauxil
   \figureopentrue\fi}
\def\FIGNUM#1{\space@ver{}\figch@ck \firstreflinetrue%
 \global\advance\figurecount by 1 \xdef#1{\the\figurecount}}
\def\FIG#1{\FIGNUM#1
   \immediate\write\figurewrite{\noexpand\refitem{#1.}}%
   \begingroup\obeyendofline\rw@start}
\def\fig{\FIGNUM\? fig.~\?%
\immediate\write\figurewrite{\noexpand\refitem{\?.}}%
\begingroup\obeyendofline\rw@start}
\def\figure{\FIGNUM\? figure~\?
   \immediate\write\figurewrite{\noexpand\refitem{\?.}}%
   \begingroup\obeyendofline\rw@start}
\def\Fig{\FIGNUM\? Fig.~\?%
\immediate\write\figurewrite{\noexpand\refitem{\?.}}%
\begingroup\obeyendofline\rw@start}
\def\Figure{\FIGNUM\? Figure~\?%
\immediate\write\figurewrite{\noexpand\refitem{\?.}}%
\begingroup\obeyendofline\rw@start}
\newcount\tablecount     \tablecount=0
\newif\iftableopen       \newwrite\tablewrite
\def\tabch@ck{\chardef\rw@write=\tablewrite \iftableopen\else
   \immediate\openout\tablewrite=tables.texauxil
   \tableopentrue\fi}
\def\TABNUM#1{\space@ver{}\tabch@ck \firstreflinetrue%
 \global\advance\tablecount by 1 \xdef#1{\the\tablecount}}
\def\TABLE#1{\TABNUM#1
   \immediate\write\tablewrite{\noexpand\refitem{#1.}}%
   \begingroup\obeyendofline\rw@start}
\def\Table{\TABNUM\? Table~\?%
\immediate\write\tablewrite{\noexpand\refitem{\?.}}%
\begingroup\obeyendofline\rw@start}
%

%
%
%
\def\masterreset{\global\pagenumber=1 \global\chapternumber=0
   \ifnum\the\equanumber<0\else \global\equanumber=0\fi
   \global\sectionnumber=0
   \global\referencecount=0 \global\figurecount=0 \global\tablecount=0 }
\def\FRONTPAGE{\ifvoid255\else\vfill\penalty-2000\fi
      \masterreset\global\frontpagetrue
      \global\lastf@@t=0 \global\footsymbolcount=0}

\def\paperstyle{\letterstylefalse\normalspace\papersize}
\def\letterstyle{\letterstyletrue\singlespace\lettersize}
\def\papersize{\hsize=35pc\vsize=48pc\hoffset=1pc\voffset=6pc
               \skip\footins=\bigskipamount}
\def\lettersize{\hsize=6.5in\vsize=8.5in\hoffset=0in\voffset=1in
   \skip\footins=\smallskipamount \multiply\skip\footins by 3 }
\paperstyle   
%
%
\def\MEMO{\letterstyle\FRONTPAGE \letterfrontheadline={\hfil}
    \line{\quad\fourteenrm FNAL MEMORANDUM\hfil\twelverm\the\date\quad}
    \medskip \memod@f}

\def\memit@m#1{\smallskip \hangafter=0 \hangindent=1in
      \Textindent{\caps #1}}
\def\memod@f{\xdef\to{\memit@m{To:}}\xdef\from{\memit@m{From:}}%
     \xdef\topic{\memit@m{Topic:}}\xdef\subject{\memit@m{Subject:}}%
     \xdef\rule{\bigskip\hrule height 1pt\bigskip}}
\memod@f
\newskip\lettertopfil
\lettertopfil = 0pt plus 1.5in minus 0pt
\newskip\letterbottomfil
\letterbottomfil = 0pt plus 2.3in minus 0pt
\newskip\spskip \setbox0\hbox{\ } \spskip=-1\wd0
\def\addressee#1{\medskip\rightline{\the\date\hskip 30pt} \bigskip
   \vskip\lettertopfil
   \ialign to\hsize{\strut ##\hfil\tabskip 0pt plus \hsize \cr #1\crcr}
   \medskip\noindent\hskip\spskip}
\newskip\signatureskip       \signatureskip=40pt
\def\signed#1{\par \penalty 9000 \bigskip \dt@pfalse
  \everycr={\noalign{\ifdt@p\vskip\signatureskip\global\dt@pfalse\fi}}
  \setbox0=\vbox{\singlespace \halign{\tabskip 0pt \strut ##\hfil\cr
   \noalign{\global\dt@ptrue}#1\crcr}}
  \line{\hskip 0.5\hsize minus 0.5\hsize \box0\hfil} \medskip }

\def\endletter{\ifnum\pagenumber=1 \vskip\letterbottomfil\supereject
\else \vfil\supereject \fi}
\newbox\letterb@x
\def\lettertext{\par\unvcopy\letterb@x\par}
\def\multiletter{\setbox\letterb@x=\vbox\bgroup
      \everypar{\vrule height 1\baselineskip depth 0pt width 0pt }
      \singlespace \topskip=\baselineskip }
\def\letterend{\par\egroup}
%
%
%
\newskip\frontpageskip
\newtoks\pubtype
\newtoks\Pubnum
\newtoks\pubnum
\newif\ifp@bblock  \p@bblocktrue
\def\PH@SR@V{\doubl@true \baselineskip=24.1pt plus 0.2pt minus 0.1pt
             \parskip= 3pt plus 2pt minus 1pt }
\def\PHYSREV{\paperstyle\PhysRevtrue\PH@SR@V}
\def\titlepage{\FRONTPAGE\paperstyle\ifPhysRev\PH@SR@V\fi
   \ifp@bblock\p@bblock\fi}
\def\nopubblock{\p@bblockfalse}

\frontpageskip=1\medskipamount plus .5fil
\pubtype={\tensl Preliminary Version}
\pubnum={0000}
\def\p@bblock{\begingroup \tabskip=\hsize minus \hsize
   \baselineskip=1.5\ht\strutbox \topspace-2\baselineskip
   \halign to\hsize{\strut ##\hfil\tabskip=0pt\crcr
   \the\Pubnum\cr \the\date\cr}\endgroup}

%
\def\title#1{\vskip\frontpageskip \titlestyle{#1} \vskip\headskip }
\def\author#1{\vskip\frontpageskip\titlestyle{\twelvecp #1}\nobreak}

\def\address#1{\par\kern 5pt\titlestyle{\twelvepoint\it #1}}
\def\andaddress{\par\kern 5pt \centerline{\sl and} \address}

\def\abstract{\vskip\frontpageskip\centerline{\fourteenrm ABSTRACT}
              \vskip\headskip }

%
%
%

\def\\{\relax\ifmmode\backslash\else$\backslash$\fi}
\def\globaleqnumbers{\relax\ifnum\the\equanumber<0%
\else\global\equanumber=-1\fi}

\def\journal#1&#2(#3){\unskip, \sl #1~\bf #2 \rm (19#3) }

\def\topspace{\hrule height 0pt depth 0pt \vskip}

\def\VEV#1{\left\langle #1\right\rangle}

\let\int=\intop         
\def\prop{\mathrel{{\mathchoice{\pr@p\scriptstyle}{\pr@p\scriptstyle}{
                \pr@p\scriptscriptstyle}{\pr@p\scriptscriptstyle} }}}
\def\pr@p#1{\setbox0=\hbox{$\cal #1 \char'103$}
   \hbox{$\cal #1 \char'117$\kern-.4\wd0\box0}}
\def\lsim{\mathrel{\mathpalette\@versim<}}
\def\gsim{\mathrel{\mathpalette\@versim>}}
\def\@versim#1#2{\lower0.2ex\vbox{\baselineskip\z@skip\lineskip\z@skip
  \lineskiplimit\z@\ialign{$\m@th#1\hfil##\hfil$\crcr#2\crcr\sim\crcr}}}
\def\leftrightarrowfill{$\m@th \mathord- \mkern-6mu
	\cleaders\hbox{$\mkern-2mu \mathord- \mkern-2mu$}\hfil
	\mkern-6mu \mathord\leftrightarrow$}
\def\lrover#1{\vbox{\ialign{##\crcr
	\leftrightarrowfill\crcr\noalign{\kern-1pt\nointerlineskip}
	$\hfil\displaystyle{#1}\hfil$\crcr}}}
%
%
%
\let\sec@nt=\sec
\def\sec{\relax\ifmmode\let\n@xt=\sec@nt\else\let\n@xt\section\fi\n@xt}
\def\obsolete#1{\message{Macro \string #1 is obsolete.}}
\def\firstsec#1{\obsolete\firstsec \section{#1}}
\def\firstsubsec#1{\obsolete\firstsubsec \subsection{#1}}
\def\thispage#1{\obsolete\thispage \global\pagenumber=#1\frontpagefalse}
\def\thischapter#1{\obsolete\thischapter \global\chapternumber=#1}
\def\nextequation#1{\obsolete\nextequation \global\equanumber=#1
   \ifnum\the\equanumber>0 \global\advance\equanumber by 1 \fi}
\def\BOXITEM{\afterassigment\B@XITEM\setbox0=}
\def\B@XITEM{\par\hangindent\wd0 \noindent\box0 }
%

%
\catcode`@=12 
\message{ by V.K.}
\everyjob{\input myphyx }
\def\etal{{\it et al.}}
%
\overfullrule=0pt

%
%
\def\undertext#1{{$\underline{\hbox{#1}}$}}

\def\simlt{\hbox{ \rlap{\raise 0.425ex\hbox{$<$}}\lower 0.65ex\hbox{$\sim$} }}
\def\simgt{\hbox{ \rlap{\raise 0.425ex\hbox{$>$}}\lower 0.65ex\hbox{$\sim$} }}

\def\that{{\hat t}}

%
%
\font\lgh=cmbx10 scaled \magstep2
\baselineskip 12pt
\voffset = -1.1truecm
\vsize=24truecm
%
%
%
\def\pac{Paczy{\'n}ski}
\def\etal{{\it et al.}}

\def\msun{M_\odot}
\def\vperp{v_{\perp}}
\def\va{v_0}
\def\te{t_e}
\def\vcirc{v_{\rm circ}}
\def\vcircR{v_{\rm circ}(R_0)}

\def\umin{u_{\rm min}}

\def\avete{\VEV{t_e}}
\def\avethat{\VEV{{\widehat t}}}

\def\Amax{A_{\rm max}}

\def\that{{\widehat t}}
%
\bigskip
\centerline{\lgh THEORY OF EXPLORING THE DARK HALO}
\medskip
\centerline{\lgh WITH MICROLENSING 1: POWER--LAW MODELS}
\bigskip
\bigskip
{\bf
\centerline {C.  Alcock$^{\ast,\dagger}$, R.A.  Allsman$^\ddagger$, \
T.S. Axelrod$^\ddagger$, D.P. Bennett$^{\ast,\dagger}$, }
\centerline{ K.H. Cook$^{\ast,\dagger}$, N.W. Evans$^{\clubsuit}$,
K.C. Freeman$^\ddagger$, K. Griest$^{\dagger,\|}$,}
\centerline {J. Jijina$^{\|}$,
M. Lehner$^{\|}$, S.L. Marshall$^{\dagger,\flat}$,
S. Perlmutter$^\dagger$,}
\centerline {B.A. Peterson$^\ddagger$, M.R. Pratt$^{\dagger,\flat}$,
P.J. Quinn$^\ddagger$, A.W. Rodgers$^\ddagger$,}
\centerline {C.W. Stubbs$^{\dagger,\flat}$, W. Sutherland$^\spadesuit$}
\centerline { (The MACHO Collaboration) }}
\vskip 0.5truein
\centerline{$^\ast$ Lawrence Livermore National Laboratory, Livermore, CA
94550}
\vskip        8pt
\centerline{$^\dagger$ Center for Particle Astrophysics, University of
California, \
Berkeley, CA 94720}
\vskip 8pt
\centerline{$^\ddagger$ Mt. Stromlo and Siding Spring Observatories,}
\centerline{Australian National University, Weston, ACT 2611, Australia}
\vskip 8pt
\centerline{$^\clubsuit$ Theoretical Physics, Department of Physics,
University of Oxford, OX1 3NP, UK}
\vskip 8pt
\centerline{$^\|$ Department of Physics, University of California, \
San Diego, CA 92039 }
\vskip 8pt
\centerline{$^\flat$ Department of Physics, University of California, \
Santa Barbara, CA 93106 }
\vskip 8pt
\centerline{$^\spadesuit$ Astrophysics, Department of Physics,
University of Oxford, OX1 3RH, UK}

\vskip 0.6truein
\centerline{Submitted to the Astrophysical Journal, 4 November, 1994}

\bigskip
\centerline{\bf Abstract}
\bigskip

If microlensing of stars by dark matter has been detected (Alcock
\etal\ 1993; Aubourg \etal\ 1993; Udalski \etal\ 1993; Alcock \etal\ 1994;
Udalski \etal\ 1994a,b), then the way is open for the development of
new methods in galactic astronomy. This series of
papers investigates what
microlensing can teach us about the structure and shape of the dark
halo. In this paper we present
formulas for the microlensing rate, optical depth and event
duration distributions for a simple set of axisymmetric
disk--halo models. The halos are based on the \lq\lq power--law models"
(Evans 1993, 1994) which have simple velocity distributions.

Using these models, we show that there is a large uncertainty in the
predicted microlensing rate because of uncertainty in the halo parameters.
For example, models which reproduce the measured galactic observables
to within their errors still differ in microlensing rate towards the
Magellanic Clouds by more than a factor of ten. We find that while
the more easily computed optical depth correlates well with
microlensing rate, the ratio of optical depth to rate can vary by a
factor of two (or greater if the disk is maximal). Comparison of
microlensing rates towards the Large and Small Magellanic Clouds
(LMC and SMC) and M31 can be used to aid determinations of the halo
flattening and rotation curve slope. For example, the ratio of
microlensing rates towards the LMC and SMC is $\sim 0.7-0.8$ for E0
halos and $\sim 1.0 - 1.2$ for E7 halos (c.f. Sackett \& Gould 1993).
Once the flattening has been established, the ratio of microlensing
rates towards M31 and the LMC may help to distinguish between models
with rising, flat or falling rotation curves. Comparison of rates
along LMC and galactic bulge lines-of-sight gives useful information
on the halo core radius, although this may not be so easy to extract
in practice. Maximal disk models provide substantially smaller halo
optical depths, shorter event durations and even larger model uncertainties.

\noindent
Subject headings:  dark matter - Galaxy: structure - gravitational lensing

\unnumberedchapters
\baselineskip  12pt

\bigskip
\centerline{\bf 1. Introduction}
\medskip
The recent detection of possible gravitational microlensing events
(Alcock \etal\ 1993; Aubourg \etal\ 1993; Udalski \etal\ 1993,
Alcock \etal\ 1994; Udalski \etal\ 1994a,b) gives hope
that at least part of the dark matter content of our galaxy is directly
accessible to observation. The dark halo, whose extent has been studied
gravitationally for many years via velocities of stars, gas, and satellites
(e.g.~Fich \& Tremaine 1991), contains at least three times
(and perhaps more than ten times) the mass of the
luminous galaxy. Its identity is
one of the major unsolved problems in astronomy (e.g.~Ashman 1992; Primack,
Seckel, \& Sadoulet 1988). While it is possible that the halo consists
mostly of exotic non-baryonic elementary particles, the idea
of \pac\ (1986) of searching for Massive Compact Halo Objects (Machos)
in the range $10^{-8}\msun$ to $10^{3}\msun$ by monitoring millions
of stars in the LMC may have borne fruit in the experimental programs.

If the Milky Way halo contains large numbers of Machos, then the
gravitational microlensing experiments now under way should have
the potential to determine the number and
distribution of Machos in the halo, as well as their mass distribution.
The next step for the microlensing experiments will be to gather more
events and then translate the number and duration of those events
into an estimate of the mass fraction $f$ of the dark halo which consists
of Machos in the relevant mass range. To accomplish this goal a model
of the dark halo is necessary. In the past simple spherical models
with flat rotation curves have been considered (\pac\ 1986; Griest 1991;
DeRujula, Jetzer \& Masso 1991; Nemiroff 1991). These have been
valuable in estimating the order--of--magnitude effects but suffer
from at least three important deficiencies:

\noindent
(1) The halo may not be spherical. N-body simulations of gravitational
collapse of collisionless dark matter generically produce axisymmetric or
triaxial halos (Quinn, \etal\ 1992;
Dubinski \& Carlberg 1991; Katz 1991). The recent
papers of Sackett \& Gould (1993) and Frieman \& Scoccimarro (1994)
have made an important start on the study of microlensing effects
in flattened halos (see also the early work of Jetzer 1991).

\noindent
(2) The effect of the galactic disk is ignored. The disk makes a
significant contribution to the local circular speed. Modeling without
proper allowance for this effect leads us to over--estimate the
gravity field and hence the mass of the dark halo.
That is, since the amount of material in the galactic halo is
set by the local circular speed, a larger contribution to this speed
by the disk, means a smaller halo is needed to explain the total circular
velocity.

\noindent
(3) It is only a simplified view of the data that permits one to
regard the rotation curve of the Milky Way as flat. In fact, even the sign of
the local gradient of the rotation law  at the sun is not known --
Fich, Blitz \& Stark (1989) estimate that it may be rising or falling by
about 30 km/sec outwards from the solar circle to a galactocentric radius
of 17 kpc.

\noindent
In this paper, we make a start towards quantifying and remedying these defects
by calculating the microlensing rate and optical depth in a set of simple,
flexible and realistic halo models. We take the power--law models
(Evans 1993, 1994, hereafter E93, E94) --  for which simple and self-consistent
distribution functions are known -- and provide an approximate method to
allow for the influence of the galactic disk. This enables calculation not
just of the optical depth, but also the event
rate and the distribution of event
durations. In this paper, the emphasis is on the uncertainties in microlensing
predictions and what can be done to reduce them. Since there are many
possible models of the dark halo consistent with current observations,
there is a substantial scatter in the predicted microlensing rate -- and
this in turn contributes to an uncertainty in the measurement of the fraction
of the halo consisting of Machos. Of course, this is directly relevant to the
difficult but important question of whether the observed microlensing rates
can rule out or support the existence of non--baryonic matter in the halo.

We show that the often calculated optical depth is a useful predictor of
the microlensing rate to a factor of $\sim 2$. For more accurate work
and to predict event durations, a distribution of Macho velocities is needed.
By considering a large range of model parameters consistent with
observations, we find the microlensing optical depth can vary by a factor
of six or more. If the disk of the Milky Way is ``maximal"  -- in the sense
that it provides almost all of the Galacto-centric acceleration --
then a much smaller halo is required. This gives an even larger spread
in predicted optical depth, rate, and event durations. An attractive
possibility is to use the measured microlensing rates toward different
sources (e.g. LMC, SMC, M31 and the galactic bulge) to determine some
of the halo and disk parameters, thereby providing a new tool for the study of
galactic structure and at the same time reducing the halo uncertainty in
the measurement of $f$. We corroborate the Sackett \& Gould (1993)
prediction that the ratio of LMC to SMC optical depth is a robust
indicator of the flattening of the dark halo -- and extend it by showing
that the ratio of microlensing rates distinguishes flatness as well.
We predict that the ratio of rates toward M31 and the LMC may be enable
us to discover whether the rotation curve is rising or falling (or
equivalently the extent of the dark halo). A comparison of the bulge
and LMC rates can provide information on the halo core radius.

The plan of the paper is as follows:
In \S 2 we review the axisymmetric models and present formulas
for the optical depth, microlensing rate, and event durations.
In \S 3 we discuss the halo parameters and their allowed range and
explain how to take into account the effect of the galactic
disk.  In \S 4 we compare optical depth and rate, and discuss the
uncertainties in microlensing rates due to uncertainties in the
parameters. In \S 5 we discuss reducing those uncertainties by
comparing results along different lines-of-sight, in \S 6 we
discuss distributions of event durations, and in \S 7 we
summarize our conclusions.


\medskip
\centerline{\bf 2. Axisymmetric models}
\medskip
The primary goal of the galactic gravitational microlensing experiments
is to determine the mass of the dark halo in Machos.  The experiments
search for Machos by monitoring millions of stars nightly in the Large
Magellanic Cloud (LMC), the Small Magellanic Cloud (SMC), the galactic
bulge, and perhaps in the future in the M31 galaxy (Crotts 1992;
Baillon \etal\ 1993). If the dark halo contains large numbers of Machos,
occasionally one passes close to the observer--star line--of--sight and
acts as a gravitational lens, causing a time-dependent magnification of
the stellar image. The resulting lightcurve is determined by only
a few quantities such as the distance $s$ from us to the Macho, Macho
mass $m$, Macho transverse
velocity $\vperp$, and impact parameter $b$.  The magnification $A(t)$ as a
function of time $t$ is given by
$$
\eqalign{
A(t) &= (u^2 +2)/[u(u^2+4)^{1/2}], \cr
u(t) &= b/R_e = (\umin^2 + \omega^2 (t-t_0)^2)^{1/2}, \cr
R_e &= (2/c) [ Gms(1-s/L)]^{1/2}, \cr}
\eqn\eqnA
$$
where the peak magnification $\Amax$ is given by inverting $\umin =
u(A_{max})$, $\omega = \vperp/R_e$, and $L$ is the distance to the
star. Experimentally, microlensing events are characterized by
the maximum magnification $\Amax$, the time of the peak $t_0$, and
the duration of the event $\that$,
where $\that = 2/\omega$.
Also used in the literature as an ``event duration" is
$\te = \that (u_T^2 - \umin^2)^{1/2}$.  This is time for which $A \ge A_T$,
with $A_T = A(u_T)$. Using $A_T=1.34$ corresponds to $u_T=1$ which is
the time inside the Einstein radius $R_e$. A more thorough discussion
can be found in in \pac\ (1986) and Griest (1991).

An observing team measures the number and duration of microlensing events.
The number of observed events is proportional to the number of stars
monitored, the duration of the experiment, the experimental efficiency,
and the rate at which microlensing occurs. The primary observables are
the optical depth $\tau$, the rate $\Gamma$ and the average duration
of the events $\avethat$. They are related by
$$
\tau = \Gamma \avete = {\pi\over 4}\Gamma \VEV{\that}.
\eqn\eqntauone
$$
If the distribution of Machos masses $n(m)$ were a delta function,
then $\Gamma$ would be $\Gamma = \Gamma_1 (m/\msun)^{-1/2}$,
where $\Gamma_1$ is the rate with $m=\msun$ and is independent of mass.
For a general normalized mass distribution $\Gamma = \eta_m \Gamma_1$, where
the mass integral is
$$
\eta_m = \int dm n(m) (m/\msun)^{-1/2}.
\eqn\eqnetam
$$
This also implies that $\avethat = \avethat_1 \eta_m^{-1}$.

The optical depth is the number of Machos inside the microlensing tube
with radius $u_TR_e(s)$ and length $L$. It depends only on the density
of Machos $\rho$
$$
\tau = \int {\rho({\bf x}) \over m} \pi u_T^2 R_e^2(s) ds.
\eqn\eqntautwo
$$
Unlike the rate or average duration, it is independent of the
Macho mass distribution.  For this reason it also contains no direct
information about the mass of the lensed objects. Note also that
since detection efficiencies depend upon the duration of events, it is
important to have models which predict durations. Now to find the rate
at which Machos enter the microlensing tube requires knowledge of the
distribution of velocities all along the tube. So the rate and the
distribution of event durations are hard to calculate because they
require the entire phase space distribution function (DF) $F({\bf v,x})$.
The differential rate is given by:
$$
d\Gamma ={1\over m} F({\bf v,x}) \cos\theta u_T R_e \vperp
d^3v dx d\alpha, \eqn\eqnrate
$$
where the angles and notation are defined in Griest (1991).

What makes the calculation particularly difficult is that the DF
cannot be prescribed arbitrarily as a Maxwellian, for instance. This is
because the Machos are collisionless, so the DF is constrained to obey
the collisionless Boltzmann equation. By Jeans' theorem, this implies
that the DF depends only on the isolating integrals of motion (see
Binney \& Tremaine 1987, p. 220). Self--consistent solutions for
distributions of velocities that build flattened halo models are
scarce. The largest known set of axisymmetric models with simple
DFs are the ``power--law galaxies" (E93, E94). These form the basis for
the exploration of microlensing in this paper, allowing us to go beyond
simple spherical models.
Note, however, that all these models are axisymmetric
and oblate, while N-body simulations
suggest that halos may well be triaxial.  Exploration of triaxial models
will be done in a future paper.

The parameters of the power--law models are:

\noindent
(1) The core radius $R_c$, which measures the scale at which the density
law begins to soften.

\noindent
(2) The flattening parameter $q$, which is the axis ratio of the concentric
equipotential spheroids, with $q=1$ representing a spherical (E0) halo
and $q \sim 0.7$ representing an ellipticity of about E6.  The ``isophotal"
ellipticity of the dark halo is a function of $q$, as well as other
parameters of the model [see E94, eq. (2.9)].

\noindent
(3) The parameter $\beta$, which determines whether the rotation curve
asymptotically rises, falls or is flat. At large distances $R$ in the
equatorial plane, the rotation velocity $\vcirc \sim R^{-\beta}$. So
$\beta=0$ corresponds to a flat rotation curve, while $\beta<0$ is a
rising rotation curve and $\beta>0$ is falling.

\noindent
(4)The solar radius $R_0$, which is the distance of the
Sun from the galactic center.

\noindent
(5) Finally, the normalization velocity $\va$, which determines the overall
depth of the potential well and hence the typical velocities of Machos
in the halo. In the limit $\beta=0$, $q=1$ and large $R$ (spherical halo
with a flat rotation curve) $\va = \vcirc$.

Using $z$ as the height above the equatorial plane, the potential of the
power--law models is
$$
\Psi = \cases{ {\displaystyle {v_0^2 R_c^\beta/\beta \over
	(R_c^2 + R^2 +z^2q^{-2})^{\beta/2}} }, & if $\beta \neq 0$,\cr
	\null & \null \cr
	\null & \null \cr
       {\displaystyle - {v_0^2\over 2}\log ( R_c^2 + R^2 + z^2q^{-2})},
       & if $\beta = 0$,\cr}
\eqn\eqnPsi
$$
and the mass density is
$$
\rho = {v_0^2 R_c^\beta \over 4 \pi G q^2}
	{ R_c^2(1+2q^2) + R^2(1-\beta q^2) + z^2(2 - (1+\beta)q^{-2})
	\over (R_c^2 + R^2 +z^2q^{-2})^{(\beta+4)/2}}.
\eqn\eqnrho
$$
The DF corresponding to this potential--density pair is
$$
F(E,L_z) =  \cases { {\displaystyle A L_z^2|E|^{4/\beta-3/2} +
		B |E|^{4/\beta -1/2} + C |E|^{2/\beta-1/2} },& if
		$\beta \neq 0$,\cr
		\null & \null \cr
		\null & \null \cr
		{\displaystyle A L_z^2 \exp (4E/v_0^2) +
		 B \exp (4E/v_0^2) + C\exp(2E/v_0^2)}, & if
		 $\beta = 0$.\cr}
\eqn\eqnF
$$
where the constants $A$, $B$, and $C$ are given in E93 and E94.
As required by Jeans' theorem, the DFs depend only on the isolating
integrals of motion, namely the relative energy per unit mass
$E = \Psi - {1\over2}v^2$, and the angular momentum per unit mass
about the symmetry axis $L_z$. The circular velocity in the
equatorial plane is
$$
\vcirc^2 = {v_0^2 R_c^\beta R^2 \over (R_c^2+R^2)^{(\beta+2)/2}}.
\eqn\eqnvcirc
$$

Note that the limit $q=1$, $\beta=0$ and $R_c=0$ recovers the standard
singular isothermal sphere used by \pac. Allowing a core radius gives
$$
\rho = {v_0^2 \over 4 \pi G} {R^2 + 3R_c^2
	\over (R^2 + R_c^2)^2}.\eqn\rhosphere
$$
This differs from the cored isothermal sphere considered by Griest (1991)
in several ways. First, the rotation curve approaches its asymptotic
value more quickly. Second, the DF given by the $q=1$, $\beta=0$ limit of
equation~\eqnF~ is self--consistent, whereas Griest (1991) assumed an
approximate Maxwellian distribution of velocities.

So far we have only modeled the dark halo. However, in the standard
model, a substantial fraction ($\sim 40 \%$) of the centripetal force at
the solar radius derives from the disk stars. This is represented by a thin
exponential disk with a scale length of $R_d=3.5$ kpc, normalized to
a surface density of $\Sigma_0 = 50 \msun {\rm pc}^{-2}$ at the solar
radius (Gilmore \etal\ 1989; Gould 1990). It is possible that the disk
of our galaxy is substantially larger than the canonical value.
(Oort 1960; Bahcall 1984; Kuijken and Gilmore 1989; Gould 1990).
Recent microlensing results (Alcock, \etal\ 1994; Udalski, \etal\ 1994)
as well as studies of the optical rotation curves of external
galaxies (Buchhorn 1992; Kent 1992) may suggest this. We consider such a
``maximal disk" by taking $\Sigma_0 = 100 \msun {\rm pc}^{-2}$. The
rotation velocity added in quadrature is thus (Freeman 1970; Binney
\& Tremaine 1987, p. 77)
$$
v_{disk}^2 = {4 \pi G \Sigma_0 h y^2}
	\left[I_0(y)K_0(y) - I_1(y)K_1(y) \right],\eqn\diskrot
$$
where $y=R/(2R_d)$, and the $I_n$ and $K_n$ are modified Bessel functions.
Note that in adding a contribution from the disk to the local circular
velocity, we have sacrificed self--consistency. Really, we should find
the DF of the power--law halo in the combined potential field of both
disk and halo -- instead, we use the DF \eqnF. As has been argued
elsewhere (Evans \& Jijina, 1994), this is an reasonable approximation
for the LMC, SMC and M31, where microlensing typically occurs at heights
above the equatorial plane of many kpc.

In this paper, our aim is to estimate the contribution of the Galactic
halo to microlensing. Of course, this is not the only possible source
of deflectors. Towards the LMC, there is the possibility of microlensing
by the LMC dark halo or disk (Gould 1993b, Sahu 1994). The optical depth
is $\sim 2.5 \times 10^{-7}$ for microlensing by LMC halo lenses and $\sim 0.09
\times 10^{-7}$ for LMC disk lenses. The Galactic halo makes a
contribution that is roughly three times greater and so is the dominant
source of lenses. However, this is not the case for lines of sight towards
M31. The optical depth is dominated by Machos in the halo and disk of M31
(Crotts 1992, Gould 1993a). Crotts (1992) estimates that the halo of our own
Galaxy contributes just $20 \%$ to the total optical depth.
Microlensing towards the Galactic bulge poses perhaps the
hardest problems of separating the contributions of different deflector
populations. Bulge stars can undergo microlensing not only by
halo Machos, but also by other bulge and disk stars (Griest \etal\ 1991,
\pac\ 1991, Kiraga \& \pac\ 1994). At Baade's Window, the optical depth
is $\sim 6.3 \times 10^{-7}$ for microlensing by bulge lenses,
$\sim 5.0 \times 10^{-7}$ for disk lenses. The dark halo only makes
an important contribution if the core radius is small.

We are now in a position to calculate the microlensing observables -- the
optical depth, rate and average duration of events. They can be found
using equations~2, 3, 4 and 7. The results are single quadratures and
readily evaluated on the computer. They are displayed in Appendix A. In
Appendix B, we give the differential microlensing rate $d\Gamma/d\that$,
where $\that$ is defined just after equation (1). The probability of
obtaining an event of duration $\that$ is just $(d\Gamma/d\that)/\Gamma$.

\medskip
\centerline{\bf 3. Range of models}
\medskip

In order to explore the scatter in microlensing observables, we build
a set of halo models which span the observationally allowed
range. The power--law galaxy models allow us to vary the flattening,
core radius and rotation law, and we consider both canonical and maximal
disks. For each parameter in the model, we therefore find the range
permitted by the observations. Then, several values of each parameter
are chosen to represent the range. We also ensure that each set of
parameters gives a model consistent with the measured Milky Way rotation
curve. So, we study the statistical properties of an ensemble of models,
each one of which is a plausible representation of the dark halo of the
Milky Way.

For the dark halo flattening, little is known. So the entire range of
flattening allowed by the power--law models is examined. This
varies between E0 or spherical ($q = 1$) and roughly E6 or E7
(depending on $\beta$). The core radius of the dark halo is also
uncertain --  Bahcall, Schmidt \& Soneria (1983) estimate $R_c$ as
2 kpc from star count data, while Caldwell \& Ostriker (1981) suggest
10 kpc. If the disk is maximal, values as large as 20 kpc are possible.
We consider values of 2 kpc, 5 kpc, 10 kpc, and 20 kpc. The parameter
$\beta$ determines the slope of asymptotic circular velocity. Between
$R_0$  and $2R_0$, the circular velocity is probably within $10-15\%$
of the I.A.U value of $220$ km/s, but whether the measured HI rotational
velocities rise or fall with $R$ depends upon estimates of the solar
position $R_0$ and the local circular speed $v_{circ}(R_0)$ (see Fich
\etal\ 1989; Jones \etal\ 1993). Beyond 20 kpc, little is known directly,
though arguments based on the kinematics of distant satellite galaxies
support the idea of a relatively flat rotation curve out to an unknown
cut--off (Fich \& Tremaine 1991). However, current theories of galaxy formation
tend to favor the alternative view that dark halos extend indefinitely,
fading into structure on larger scales. So, we do not consider a halo
cut--off in this paper -- it would add yet another poorly known parameter
to our model. We investigate power--law halos with $\beta =$ -0.2, 0, and
0.2. These correspond to rotation curves which rise by $\sim$15\%, are flat,
or fall by $\sim$15\% between the solar radius and twice the solar
radius, depending a little upon $R_c$.

The value of the solar radius $R_0$ has been reviewed by Reid (1989).
He shows that most recent determinations lie between 7 kpc and 9 kpc,
with 7.7 kpc being his preferred value. This differs considerably from
the IAU value of 8.5 kpc (Kerr \& Lynden--Bell 1986). We examine
the values $R_0 = $ 7, 8, and 9 kpc. Finally, perhaps the single
most important parameter is the normalization velocity $v_0$. Given
our fixed disk contribution to the total rotation law, the parameter
$v_0$ is now specified once we settle upon a choice for $\vcircR$.
Merrifield (1992) estimates $\vcircR = 200 \pm 10$ km/s, Fich, Blitz,
\& Stark (1989) give $\vcircR = 220 \pm 30$ km/s, while
Rohlfs \etal\ (1986) give values between 170 km/s and 200 km/s between
$R_0 = 6$ kpc and $R_0=16$ kpc. For our ensemble of models, we impose
the constraint that the total circular velocity lies between 180 km/s
and 250 km/s at $R_0$ and $2R_0$. Note that the IAU value is 220 km/s
(Kerr \& Lynden-Bell 1986). We also investigated a more restricted
ensemble of models with $190 \leq \vcircR \leq 230$ km/s.  We
find all our results also hold for this more restricted ensemble.

\medskip
\centerline{\bf 4. Uncertainties in the Rates}
\medskip

First, let us consider the difference caused by using the optical
depth instead of the microlensing event rate. The optical depth to
microlensing is the mean number of Machos in the microlensing tube;
that is the number of microlensing events taking place at a given moment.
It is easy to calculate since it is independent of lens mass
and velocity, and only requires knowledge of the density distribution
$\rho({\bf x})$. For this reason, it is the most widely estimated
quantity. But how well does it trace the microlensing rate?

We are able to answer this question since both the rate $\Gamma$
(equation~A1) and the optical depth $\tau$ (equation A6) are known for
the power--law models. One way to test this is to plot $\avethat$, which
is the ratio of optical depth $\tau$ and $\Gamma$,
$\avethat = {4\over \pi}\tau/\Gamma$,
for many different models. The average duration $\avethat$ is a constant
if $\tau$ and $\Gamma$ are well--correlated. In Fig.~1, we show
histograms of $\avethat$ for microlensing towards the LMC, SMC and M31 for
our ensemble of models. Figs.~1a--c demonstrate that $\avethat$ tends
to vary by more than a factor of two between models. Figs.~1d--f show
an even larger spread for maximal disk models. Figs.~2a-f show this
another way by plotting the rate vs the optical depth for the set of models.
These plots show that $\avethat$ is indeed much less model dependent
than either $\tau$ or $\Gamma$.  While the rate and $\tau$ vary
by more than a factor of ten in these plots, their ratio varies
only $\sim 2$ for a canonical disk.
In fact, we note that the line $\Gamma \propto \tau^{3/2}$ is a fairly
good fit to all the models we have considered.
\foot{
To the extent that the relation $\Gamma = a \eta_m f^{-1/2} \tau^{3/2}$
holds, where $a$ is a constant from theory, we have that
$a = f^{-1/2} \eta_m^{-1} (4/\pi)^{3/2} \avethat^{-3/2} \Gamma^{-1/2}$
is independent of the model parameters.  Thus, if the macho fraction $f$
were known, one could extract the mass integral $\eta_m$ from observables
$\eta_m \approx f^{1/2} a^{-1} (4/\pi)^{3/2} N_{eff} E^{1/2}
(\sum \that_i/\epsilon_i)^{-3/2}$, where $E$ is the total exposure,
$N_{eff} = \sum \epsilon_i^{-1}$,  $\epsilon_i$ is the efficiency at which
events of duration $\that_i$ are recovered, and the sums go from 1 to
the number of observed microlensing events.  For LMC microlensing
in our set of models
we find $a \approx 3850 \pm 260{\rm yr}^{-1}$.
The physical basis for this relationship may simply be that
the optical depth is proportional to the mass along the line of
sight $\propto v_c^2$,
and the rate is proportional to the optical depth times $v_c$.
}
Thus the large scatter
in $\avethat$ seen in the maximal disk histograms is mostly just due to
the large scatter in rate.  (The rate varies more than the optical depth.)
In all the plots we use $m=\msun$, but for an arbitrary mass distribution
just scale $\Gamma$ by $eta_m$ and $\avethat$ by $\eta_m^{-1}$.
Keep in mind that a given experiment can produce only one point
in the $\Gamma$, $\tau$ plane and that the primary use of a measurement
will be to find $f$, the Macho fraction.
We see that for approximate
work, the optical depth does a reasonable job of predicting the rate.
But for more detailed work, especially when efficiencies
are involved, the difference between rate and optical depth
should be kept in mind. We also note that the predicted distribution of event
durations is found as a differential rate (Appendix B).

Next let us turn to scatter in the predicted microlensing rate
caused by uncertainties in the halo parameters. Fig.~2 shows that for all
lines of sights, there is a scatter in the rate of more than a factor
of ten for a canonical disk. For the LMC, the models with the smallest
rate have spherical halos with small core radii, falling rotation curves,
and small values of $v_0$, while the models with the largest rates have
either spherical or flattened halos, but large core radii, rising
rotation curves, and large values of $v_0$. This is as expected, since
any model which puts more mass at a large distance in the direction of
the LMC will have a larger microlensing rate, and a larger optical depth.
This is shown in Fig.~3 in which we plot the optical depth against the
rotation velocity at $r=50$ kpc. The correlation between $\tau$ and
$v_c^2(r=50{\rm\ kpc})$, while not perfect, is quite good. Note that the mean
value of the rotation velocity at $R_0$ is nearly independent of the
microlensing rate.  Figs.~2d--f and Fig.~3c-d show the case of a maximal
disk.  Here we see that the rate and optical depth can be considerably smaller
than for a canonical disk.  Also there is a
variation between models of several orders of magnitude. This is as
expected since in these models the disk is the main contributor to the
rotation curve at the solar distance. Thus a smaller enclosed halo mass
is required
to match observations, and the halo parameters are poorly constrained.

The halo may only consist of a fraction $f$ of baryonic matter
in the form of Machos. Thus, a factor of more than ten uncertainty in the
predicted rate caused by the poorly determined halo parameters
makes it difficult to determine the allowed amount of non--baryonic
dark matter. It is clearly essential to reduce the uncertainty.

\medskip
\centerline{\bf 5. Reducing Model Uncertainties}
\medskip
The primary way of reducing the model uncertainties in the microlensing
observables is to determine the halo parameters. Even within the restricted
framework of the power--law galaxy models, if $\beta$, $v_0$, $q$, $R_0$
and $R_c$ are known, there is still uncertainty in the rate. This is
because the DFs equation~\eqnF~ are the simplest consistent with the potential
and the density, but are certainly not unique. There are still further
DFs that depend on non--classical third integrals of motion and
generate anisotropic velocity distributions. Note, too, that even though
our models give a plausible representation of the Milky Way, there
certainly exist other alternatives (see e.g., Frieman \& Scoccimarro 1994;
Gates \& Turner 1993; Giudice, Mollerach \& Roulet 1994) with
different lensing properties.  And of course, the size of the disk
plays a crucial role.

One obvious way to determine halo parameters is to use conventional
astronomical techniques -- observations of stars, gas and satellites
-- to fix the solar radius and circular speed more accurately.
For example, fixing the solar radius at 8 kpc, and demanding
$\vcirc=220$ km/s  $\pm 5$\%  between 8 and 16 kpc reduces the spread
in microlensing rates toward the LMC from more than a factor of ten to
a little more than a factor of two (for the canonical disk). Uncertainties
in $\tau$ and $\avethat$ are reduced similarly. A better determination of
the halo core radius by stellar observations would also be important.

However, it is also possible to use the microlensing experiments themselves
to determine the halo parameters and reduce the model uncertainty. The
basic idea is to exploit the fact that there are at least four
viable lines-of-sight out of the Milky Way in which to measure
the microlensing rate and average event duration.  Each line-of-sight
(LMC, SMC, M31 and the bulge) offers a different ``pencil beam"
through the dark halo, and so by comparing the rates, optical depths,
and average durations among the different lines-of-sight information
concerning the halo shape can be gained. Several of the parameters,
such as flattening $q$ and asymptotic slope of the rotation law $\beta$,
may best be determined this way. So microlensing gives us a new probe
of the density and velocity structure of the dark halo. This is in
addition to information on the size of the disk gained via microlensing.

For instance, a scatter plot of the ratio of LMC and SMC rates vs the LMC
and SMC average durations is shown in Fig.~4a. The models clearly fall
into two distinct groups. Those models marked with a circle all have round
halos (E0), while those with a square are flattened to roughly E6.
Thus the ratio of LMC rate to SMC rate is a excellent indicator
of halo flattening.  This effect was first discovered  -- using
optical depth rather than microlensing rate -- by Sackett \& Gould (1993).
Frieman \& Scoccimarro (1994) have recently cautioned that the robustness
of this diagnostic may be lost if the halo is tilted with respect to the
disc -- although such a configuration cannot be a long--lasting equilibrium.
So, the halo flattening can probably be determined if enough events
are found to allow accurate measurement of the SMC microlensing rate.
Figs.~4b and 4c show the rate ratio for M31/LMC, and M31/SMC. While
separation of flattened models is still evident, one sees from the
figures that it is the LMC and SMC position relative to the halo axis
of symmetry that make the measurement of the flattening so easy. Note
again, that in an experiment one measures only one LMC rate (and optical
depth) and one SMC rate, and so gets only one point in any of these
scatter plots. It is also interesting to observe from Fig.~4a that the
model uncertainties in the LMC/SMC rate ratio are much greater for
flattened halos than for spherical halos. The case of a maximal disk
is not shown, since it looks almost identical to Fig.~4.

Can we use microlensing to determine whether the halo has a rising
or falling rotation curve? The LMC and SMC are at nearly the same
distances (50 and 60 kpc), so it is natural to expect the ratio of
M31 to LMC microlensing to be the most useful discriminant. Note that
rate ratios are convenient to use, because the magnitude of any rate
always contains the unknown parameter $f$. In Fig.~5 we plot the
M31/LMC rate ratio vs the LMC rate for the set of models above,
with triangles for $\beta=0.2$ (falling rotation curve), circles
for $\beta=0$ (asymptotically flat rotation curves), and stars
for $\beta=-0.2$ (rising rotation curve). In Fig.~5a, all models
are plotted, while in Fig.~5b and Fig.~5c only models with spherical ($q=1$)
and flattened ($q=0.71$ or $q=0.78$) halos respectively are shown.
In Fig.~5a some separation of models with different values of $\beta$
is evident but there is substantial ambiguity, which would make
a direct estimate of $\beta$ using this method difficult. However, suppose
that we have already determined the halo flattening by use of the ratio
$\Gamma_{LMC}/\Gamma_{SMC}$. Then, as shown in Figs.~5b and 5c for a
canonical disk, a fairly clear separation of rising, falling, and flat
rotation curve parameter can be accomplished. Thus, the ambiguity seen
in Fig.~5a is largely removed when models with different flattenings
are plotted separately. The exception is some overlap between models with
$R_c=2$ kpc and $R_c=20$ kpc and different values of $\beta$. This
ambiguity is probably removable as discussed below. The case of a maximal
disk is not displayed, as it is very similar. So, the asymptotic form of
the rotation law, or equivalently $\beta$, can probably be determined
from the M31/LMC rate ratio once $q$ is known. Keep in mind, however,
the caveats mentioned in \S3 concerning our M31 rate calculation, which
may result in corrections which modify this effect. If halo microlensing
can be distinguished from M31 microlensing, a measurement of $\beta$
should then be possible.

Next, can we determine the halo core radius $R_c$?  The parameter
$R_c$ affects mainly the inner portion of the halo and overall
normalization of the halo mass. This overall normalization is mixed in
with $v_0$ and $f$, and so the best hope in determining $R_c$ is
probably a comparison of the bulge with a more distance source such as
the LMC. Here we have the problems mentioned in \S3 concerning bulge
microlensing; our modelling of the distribution of velocities is
not adequate along the disk. But, the optical depth is independent of
the velocities and will give some indication of the rate. Even so,
our calculations do not give the total optical depth towards the bulge,
merely the contribution of the optical depth from the halo.

In Fig.~6, we plot the LMC/bulge optical depth ratio vs the bulge
optical depth, where triangles indicate $R_c=2$ kpc, boxes indicate
$R_c=5$ kpc, and stars indicate $R_c=10$ kpc.  A reasonably clean
separation is obtained when this ratio is plotted for all the models
(not shown). In Fig.~6, this separation is made clear--cut, if one
supposes $\beta$ and $q$ have already been measured by the methods above.
The $R_c=20$ kpc models have an LMC/bulge ratio of greater than 10,
and are very easily distinguished even with no prior knowledge
of $\beta$ and $q$.  (They fall off the top of the plots in Fig.~6).
Even if $\beta$ and $q$ are not known, the separation is quite good
if the value of the solar radius $R_0$ is held fixed. So, a better
determination of $R_0$ by non-microlensing means can allow a clearer
separation of the effect of the halo core radius. The case of a
maximal disk is not shown since it gives very similar results.

\medskip
\centerline{\bf 6. Distribution of Event Durations}
\medskip
Since the duration of a microlensing event is proportional to
the Einstein radius ($\propto m^{1/2})$, the duration
of an event gives information about the mass of lens which caused it.
In trying to understand the nature of the objects responsible for the
observed microlensing, this is important information. But the duration
also depends upon the unknown lens velocity and distance. Thus, a given
mass Macho can cause a wide distribution of event durations. This
distribution must be used statistically to infer probable masses
from observed durations. Using the DF's (equation~\eqnF), the
distribution of event durations can be found. The formula and
definitions are given in Appendix B. In Fig.~7, we show several
$\that$ distributions.  One sees that different halo parameters give
quite different distributions. It is the average of these
distributions $\VEV{\that}$ that is shown in
the histograms in Fig.~1.  Fig.~7 shows that, as expected, uncertainty
in the halo model will lead to additional uncertainty in determining
the masses of the lensing objects.
The curves labeled (a), (b), and (c) are canonical disk cases with
various choices of halo parameters, while curve (d) shows a maximal
disk example.
We also note that the scaling introduced in Griest (1991) works fairly well
for models we considered.  That is, by scaling the $\that$ axis
by $\avethat^{-1}$, and the $d\Gamma/d\that$ axis by $\avethat$,
all the curves are found to lie roughly on top of each other.
This means that for power law galaxy models along a given line-of-sight,
the shape of the distribution is much more model
independent than peak value.

In a future paper we plan to explore further
the information that can be extracted from event duration distributions,
and include other possibilities such triaxiality, streaming motion, etc.

\medskip
\centerline{\bf 7. Conclusions}
\medskip

This paper has shown how to exploit the power--law galaxy models
(E93, E94) as simple, flexible and realistic representations of
the dark halo. These models have the advantage of simple and analytic
phase space distribution functions and therefore permit accurate
calculation of the optical depth, microlensing rate and average event
duration. We provide formulae for these quantities as a function of the
halo parameters and source distance and direction (Appendix A). The
distribution of event timescales is presented in Appendix B.
We apply our formulae to study microlensing towards the Large and
Small Magellanic Clouds (LMC and SMC), the galactic bulge, and the M31
disk galaxy. We find that:

\noindent
(1) For a canonical disk, the optical depth is a reasonable
indicator of the microlensing rate to within a factor of two.
This is important, because the optical depth is much easier
to calculate than the rate and probably will continue to be widely
used by investigators.
For more accurate
work, as well as for derivations of the distribution of durations,
galaxy modeling with distribution functions is crucial.
For a maximal disk the agreement between optical depth and rate is
less robust, though the relation $\Gamma \propto m^{-1/2} \tau^{3/2}$
seems to hold.

\noindent
(2) The evaluation of the fraction $f$ of the halo consisting of Machos
is hampered by the uncertainties in the galactic constants, such
as the shape of the rotation law and the flattening of the dark halo.
For a realistic set of halo models, we found rates toward the LMC and
SMC can vary by more than a factor of ten from model to model for
a canonical disk, and by several orders of magnitude for a maximal disk.
Left unaddressed, this model uncertainty will thwart accurate
determination of $f$.

\noindent
(3) An attractive way of reducing the uncertainty -- which simultaneously
opens up a new method in galactic astronomy -- is to use microlensing
to explore the shape and structure of the dark halo. This has also
been realised by Sackett \& Gould (1993), who showed that the ratios
of optical depth towards the LMC and SMC is a robust indicator of the
flattening of the dark halo. We confirm this result by showing that the
ratios of the event rates also distinguish
flatness. In particular, the ratio of microlensing rates towards
the LMC and SMC is $\sim 0.7-0.8$ for E0 halos and $\sim 1.0 -1.2$
for E7 halos. This is true for both canonical and maximal disk models.
Once the flattening has been established, the asymptotic
slope of the rotation curve $\beta$ might be determined using the
M31/LMC rate ratio.  The LMC/bulge ratio contains important information
on the halo core radius. We caution that this may not be easy
to extract, as the dark halo is probably not the dominant source of lenses
towards the bulge.

In summary, the discovery of a dark halo consisting of a significant
fraction of Machos is only the starting point for an exploration of the
halo characteristics which microlensing can help determine.

\medskip
\centerline{\bf Acknowledgements}
KG thanks A.Gould, D.A.Merritt, and D.N.Spergel for help in the early
stages of this project.
KG acknowledges a DOE OJI grant, and KG and CWS
thank the Sloan Foundation for their support.
Work performed at LLNL is supported by the DOE under contract W7405-ENG-48.
Work performed by the Center for Particle Astrophysics on the UC campuses
is supported in part by the Office of Science and Technology Centers of
NSF under cooperative agreement AST-8809616.
Work performed at MSSSO is supported by the Bilateral Science
and Technology Program of the Australian Department of Industry, Technology
and Regional Development.

\medskip
\centerline{\bf Appendix A}

In this appendix, we give the formulae for the microlensing rate and
optical depth for the general flattened halo model described in the
text (equations~4--7). The total rate $\Gamma$ of microlensing in a
power--law halo with model parameters $\beta, v_0, R_c, R_0$, and $q$ is
$$\eqalign{\Gamma =& {C_0u_T\over \sqrt{2\pi M/M_\odot}}{v_0^3R_c^{3\beta /2}
(\beta +2)(1-q^2) \over 2cq^2\sqrt{-\beta}L^{1/2 + 3\beta/2} }{\Gamma
(n_\beta)\over \Gamma(d_\beta)}I_1\cr
+& {C_0u_T\over \sqrt{2\pi M/M_\odot}}{v_0^3R_c^{2+ 3\beta/2}(\beta +2)\over
cq^2\sqrt{-\beta} L^{5/2 + 3\beta/2}}{\Gamma(n_\beta)\over
\Gamma(d_\beta)}I_2\cr
+& {C_0u_T\over \sqrt{2\pi M/M_\odot}}{v_0^3R_c^{3\beta /2}(2 - {1+
\beta \over q^2})\over c\sqrt{-\beta}L^{1/2 + 3\beta /2}}{\Gamma
(n_\beta - 2/ |\beta|)\over \Gamma (d_\beta - 2/|\beta|)}I_3.\cr}
\eqno({\rm A}1)
$$
Here, $C_0 = 1/\sqrt{G M_\odot}$, $\Gamma(x)$ is the gamma
function, and the integrals $I_i$ are
$$\eqalign{I_1 =& \int_0^1 {ds \sqrt{s(1-s)} (A's^2 + B's + C')\over
(D's^2 + Es' +F')^{2 + 3\beta /2}}\cr
I_2 =& \int_0^1 {ds \sqrt{s(1-s)}\over (D's^2 + Es' +F')^{2 + 3\beta/4}}\cr
I_3 =& \int_0^1 {ds \sqrt{s(1-s)}\over (D's^2 + Es' +F')^{1 +
3\beta /4}}.\cr}
\eqno({\rm A}2)
$$
with
$$\eqalign{A' =& 3\cos^2 b,\quad B' = -6R_0\cos b \cos \ell /L,\cr
C'=& 2R_0^2/L^2 + R_0^2\cos^2 \ell /L^2 + R_0^2 \sin^2 \ell \sin^2
b/L^2,\cr
D'=&\cos^2 b + q^{-2}\sin^2 b,\quad E' = -2R_0\cos b\cos \ell/L\cr
F'=& (R_c^2 + R_0^2)/L^2.\cr}
\eqno({\rm A}3)
$$
The quantities $b,l$ are the galactic coordinates of the source star,
$L$ is the source distance, $G$ is Newton's constant, and $c$ is the speed
of light.  The constants $n_\beta$ and $d_\beta$ have a different form
according to whether $\beta$ is positive or negative
$$n_\beta = \cases{ {\displaystyle {-4\over \beta} - {3\over 2}},&
if $\beta <0$,\cr
\null & \null\cr
{\displaystyle {4\over \beta} +2},& if $\beta >0$,\cr}
\eqno({\rm A}4)
$$
$$d_\beta = \cases{ {\displaystyle {-4\over \beta} - 1},&
if $\beta <0$,\cr
\null & \null\cr
{\displaystyle {4\over \beta} + {5\over 2}},& if $\beta >0$,\cr}
\eqno({\rm A}5)
$$
In the limit $\beta \rightarrow 0$ (the case of an asymptotically flat
rotation curve), the expression for the rate follows from the above by
systematic use of the formula $\Gamma ( x+ 1/2) / \Gamma (x)
\rightarrow \sqrt{x}$ as $x \rightarrow \infty$.
The optical depth $\tau$ is
$$\tau = {v_0^2R_c^\beta u_T^2\over c^2 q^2 L^\beta}\int_0^1
{s(1-s)(A''s^2 + B''s +C'')ds \over (D's^2 + E's +F')^{(\beta+4)/2}}.
\eqno({\rm A}6)
$$
where
$$\eqalign{&
A'' = (1- \beta q^2)\cos^2 b + (2 - (1+\beta)q^{-2})\sin^2 b,\cr
&B'' = -2(1-\beta q^2)R_0\cos b\cos \ell/L ,\cr
&C'' = (R_c^2(1+2q^2) + R_0^2(1-\beta q^2))/L^2.\cr}
\eqno({\rm A}7)
$$
The quadratures are straightforward to evaluate on the computer.
\bigskip\bigskip
\centerline{\bf Appendix B}
\medskip
The distribution of event durations is important for finding the mass
of the lensing objects.  It is given by the normalized differential
microlensing rate $(d\Gamma /d\that)/\Gamma$, where $\that =
2 R_e/v_\perp$, and $v_\perp$ is the speed of the Macho perpendicular
to the line-of-sight.  The time the Macho spends inside
the Einstein radius, $t_e = (u_T^2-u_{min}^2)^{1/2}\that$, where $u_{min}$ is
defined in equation~\eqnA, and $u_T=1$. The average duration is
related to the average $\that$ by $\avete = {\pi\over 4}\VEV{\that}$.
In many cases it is advantageous to use distributions in $\that$, since
they are independent of the amplifications.

For the model described in the text, we find:
$$\eqalign{{d\Gamma\over d{\hat t}}&=8\,{u_T\over \pi c^2}\,\left(L^6\over
R_C^4 {\hat t}^4\right)\,(\beta +2)\,|\beta|^{1+4/\beta}\,
(q^{-2}-1)\,\space [a_1 G'\, J_{1}\space + a_2\space H'\,J_2]\cr
&+\,8\, {u_T\over \pi c^2}\,\left(L^4\over R_c^2{\hat t}^4\right)\,
\,{|\beta|^{1+4/\beta} (\beta +2)\over q^2}\,a_1\,J_1\cr
&+\,8\, {u_T\over \pi c^2}\,\left(L^4\over R_c^2{\hat t}^4
\right)\, |\beta|^{1+2/\beta} \,(2-q^{-2}(1+\beta))\,a_3\,J_3}
\eqno({\rm B}1)$$
where,
$$\eqalign{
a_1 &= \cases{ {-1-{4\over \beta}},&$\beta<0$\cr
1+{4\over \beta}, &$\beta>0$,}\cr
a_2 &= {4 (\beta +4)\over \beta^2},\cr
a_3&=\cases{ {-1-{2\over \beta}}, &$\beta<0$\cr
1+{2\over \beta}, &$\beta>0$,\cr}}\eqno({\rm B}2)$$
and the integrals $J_i$ are
$$\eqalign{
J_1 &=\int\,ds\,s^2(1-s)^2\,\left | {K'\over {g_1(D's^2+E's+F')^{\beta/2}}}
\, -\, {H'\over g_1}\,s(1-s)\right |^{4/\beta},\cr
J_2&=\int\,ds\,s^3(1-s)^3\, \left[{A'\over 3}s^2+{B'\over 3}s +
(C'-2\left(R_0\over L\right)^2)\right]\,\times \cr
&\qquad\qquad \left |{K'\over {g_2(D's^2+E's+F')^{\beta/2}}}\, -\,
{H'\over g_2}\,s(1-s)\right|^{4/\beta -1},\cr
J_3&=\int\,ds\,s^2(1-s)^2\,\left | {K'\over {g_3 (D's^2+E's+F')^{\beta/2}}}
\,-\, {H'\over g_3}\,s(1-s)\right|^{2/\beta}.}\eqno({\rm B}3)$$
If $\beta < 0$, the integrals are evaluated over the interval $[0,1]$. If
$\beta > 0$, then we must restrict the domain of integration by
$$ {\hat t}^2 \ge {8\beta L\,s(1-s)\,(m/M_{\odot})\,
(D's^2+E's+F')^{\beta/2}\over (v_0 c C_0)^2 (R_C/L)^\beta}.\eqno({\rm
B}4)$$

\noindent The constants $A'$, $B'$, $C'$, $D'$ and $E'$ are given in
Appendix A. The additional constants are
$$\eqalign{G'&=\left({R_0\over L}\,\cos b\,\sin l\right)^2,\qquad\qquad
H'={8\over L\,(c C_0)^2}\,{m\over M_{\odot}},\cr
K'&=\left(v_0 {\hat t}\over L\right)^2\,{1\over \beta}\,
\left(R_C\over L\right)^\beta,\qquad\qquad
g_1=H'^{(-\beta/4)}\,\left(v_0 {\hat t}\over L\right)^2,\cr
g_2&=H'^{\left(\beta \over \beta -4\right)}\,
\left(L\over v_0 {\hat t}\right)^{8\over \beta -4},\qquad\qquad
g_3=H'^{(-\beta/2)}\,\left(v_0 {\hat t}\over L\right)^2.} \eqno({\rm B}5)$$

\bigskip\bigskip
\centerline{\bf References}
\medskip
\def\refind{\noindent\hangindent=1.5pc\hangafter=1}

\refind
Alcock, C. \etal, 1993, Nature, 365, 621

\refind
Alcock, C. \etal, 1994, ApJ, in press

\refind
Ashman, K. 1992, PASP, 104, 1109

\refind
Aubourg, E. \etal, 1993, Nature, 365, 623

\refind
Baillon, P., Bouquet, A., Giraud-Heraud, Y., \& Kaplan, J. 1993,
A\&A, 277, 1

\refind
Bahcall, J., Schmidt, M., \& Soneira, R., 1983, 265, 730

\refind
Bahcall, J., 1984, \sl Ap. J., \bf 287, \rm 926.

\refind
Binney, J. \& Tremaine, S. 1987, Galactic Dynamics (Princeton University
Press, Princeton)

\refind
Buchhorn, M. 1992, PhD Thesis, Australin National University

\refind
Caldwell, J.A.R., \& Ostriker, J.P. 1981, ApJ, 251, 61

\refind
Crotts, A.P.S. 1992, ApJ, 399, L43

\refind
DeRujula, A., Jetzer, Ph., \& Masso, E. 1991, MNRAS, 250, 348

\refind
Dubinski, J. \& Carlberg, R., 1991, ApJ, 378, 496

\refind
Evans, N.W. 1993, MNRAS 260, 191 (E93)

\refind
Evans, N.W. 1994, MNRAS, 267, 333 (E94)

\refind
Evans, N.W. \& Jijina, J. 1994, MNRAS, 267, L21

\refind
Fich, M. \& Tremaine, S. 1991, ARAA, 29, 409

\refind
Fich, M., Blitz, L. \& Stark, A.A. 1989, ApJ, 342, 272

\refind
Freeman, K.C. 1977, ApJ, 160, 811.

\refind
Frieman, J. \& Scoccimarro, R. 1994, ApJ, 431, L23.

\refind
Gates, E. \& Turner, M.S. 1993, preprint FERMILAB-Pub-93/357-A

\refind
Giudice, G.F., Mollerach, S., \& Roulet, E. 1994, Phys. Rev. D, in press

\refind
Gilmore, G, Wyse, R.F.G., \& Kuijken, K. 1989, ARAA, 72, 555

\refind
Gould, A. 1990, MNRAS, 244, 25

\refind
Gould, A. 1992, ApJ, 392, 442

\refind
Gould, A. 1993a, private communication

\refind
Gould, A. 1993b, ApJ, 404, 451

\refind
Griest, K. 1991, ApJ, 366, 412

\refind
Griest, K. \etal\ 1991, ApJ, 372, L79

\refind
Jetzer, Ph. 1991,  in ``Atti del Colloquio de Mathematica", vol 7,
ed. CERFIM, Locarno

\refind
Jetzer, Ph. \& Masso E., 1994, Phys. Lett. B., 323, 347

\refind
Jones, B., \etal\ 1993, Lick preprint

\refind
Katz, N. 1991, ApJ, 368, 325

\refind
Kent, S.M. 1992, ApJ, 387, 181

\refind
Kerr, F.J. \& Lynden-Bell, D. 1986, MNRAS, 221, 1023

\refind
Kiraga, M. \& Paczy\'nski, B., 1994 ApJ, 430, L101

\refind
Nemiroff, R.J. 1989, ApJ, 341, 579

\refind
Nemiroff, R.J. 1991, A\&A, 247, 73

\refind
Merrifield, M.R. 1992, AJ, 103, 1552

\refind
Oort, J.~H. 1960, \sl Bull. Astr. Inst. Netherlands, \bf 6, \rm 249.

\refind
\pac, B. 1986, ApJ, 304, 1

\refind
\pac, B. 1991, ApJ, 371, L63

\refind
Primack, J.~R., Seckel, D., \& Sadoulet, B. 1988, Ann. Rev. Nucl.
Part. Sci., 38, 751

\refind
Reid, N.J. 1989, in The Center of the Galaxy, IAU Symposium No. 136.
	ed. Morris, M. (Kluwer, Dordrecht)

\refind
Rohlfs, K., Chini, R., Wink, J.E., \& Bohme, R. 1986, A\&A, 158, 181

\refind
Sackett, P.~D. \& Gould, A. 1993, ApJ 419, 648

\refind
Sahu, K., 1994, Nature, 370, 275

\refind
Udalski, A., \etal, 1993, Acta Astron, 43, 289

\refind
Udalski, A., \etal, 1994a, ApJ, 426, L69

\refind
Udalski, A., \etal, 1994b, Acta Astron, 44, 165

\refind
Warren, M.S., \etal,  1992, ApJ 339, 405


\vfill
\eject
\centerline{\bf Figure Captions}
\medskip
\noindent
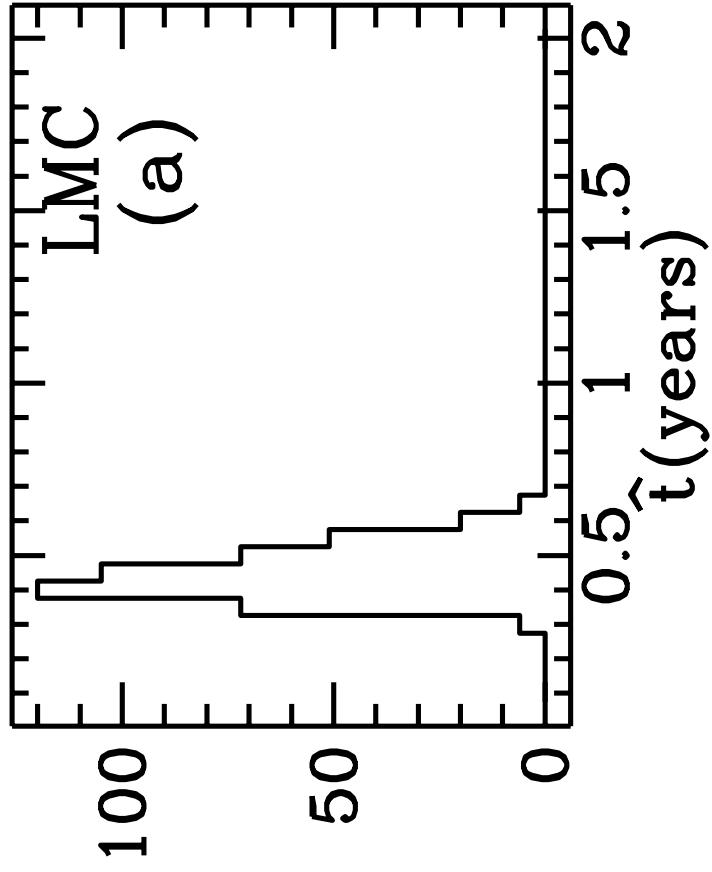
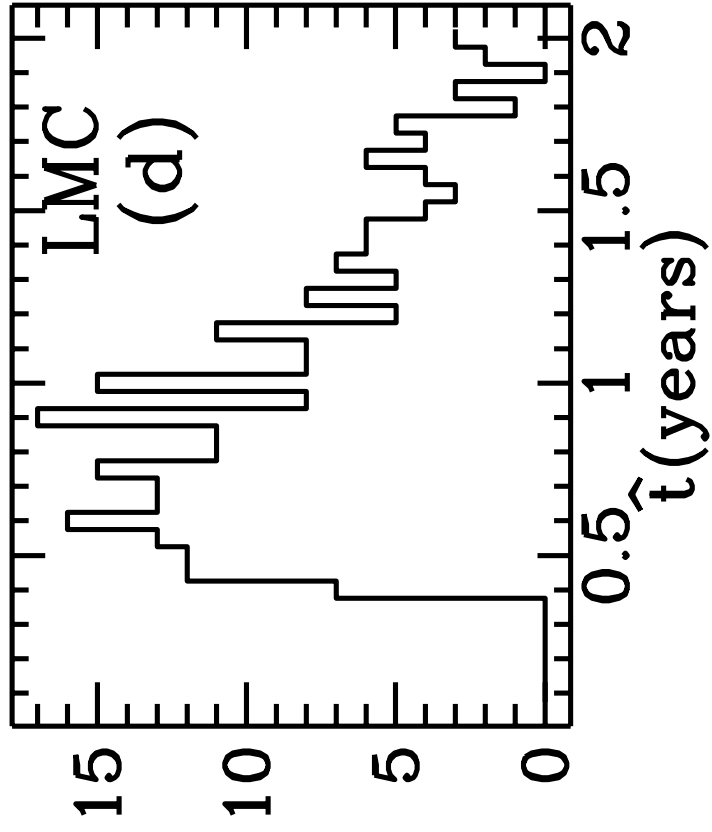
{\bf Figure 1:}
Histograms of the average duration $\avethat =
{4\over\pi}\tau/\Gamma$ for the ensemble
of halo models discussed in the text.  Part (a) is for the LMC,
(b) is for the SMC, and (c) is for M31.  If optical depth tracked microlensing
rate perfectly each histogram would be a delta function.
Parts (d)-(f) are the same for a maximal disk model.
Note all plots are for $m=1\msun$;  scale by $\eta_m^{-1}$ for other masses
(Equation \eqnetam).
\medskip

\noindent
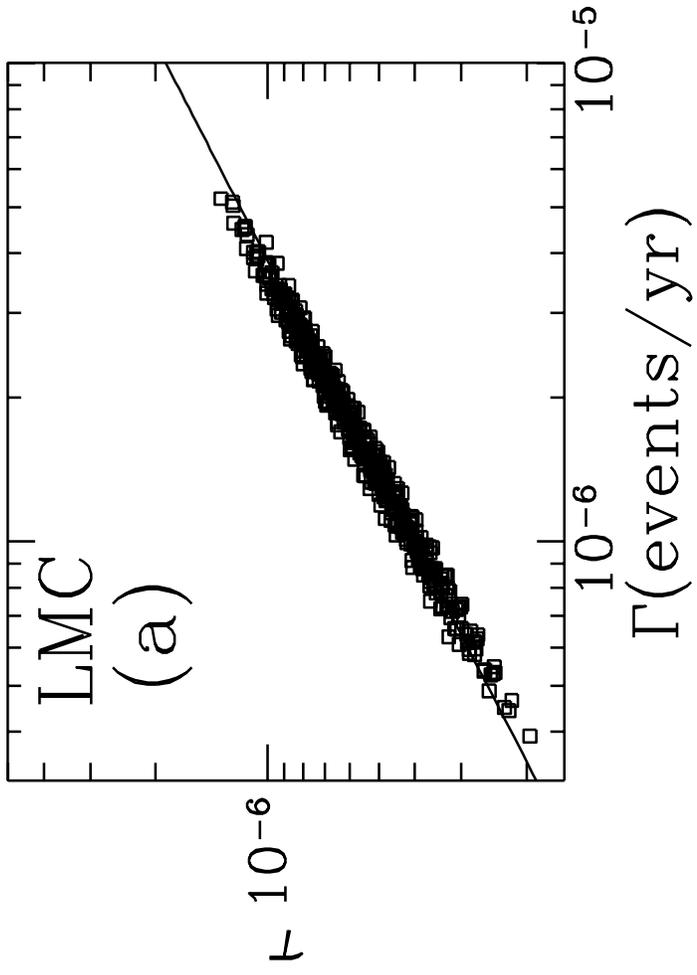
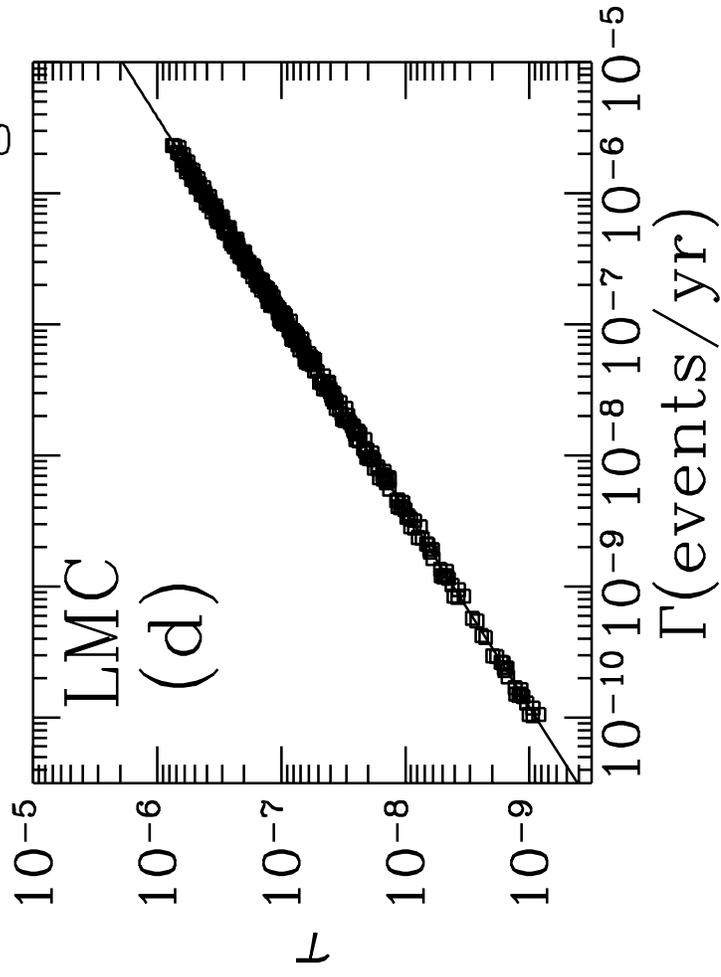
{\bf Figure 2:}
Scatter plots of microlensing rate vs optical depth for the ensemble of
models discussed in the text.
Part (a) is for the LMC, (b) is for the SMC, and (c) is for M31.
Each point represent a consistent model of the dark halo.
Parts (d)-(f) are the same for a maximal disk model.  All event
rates scale $\Gamma \propto \eta_m$.
\medskip

\noindent
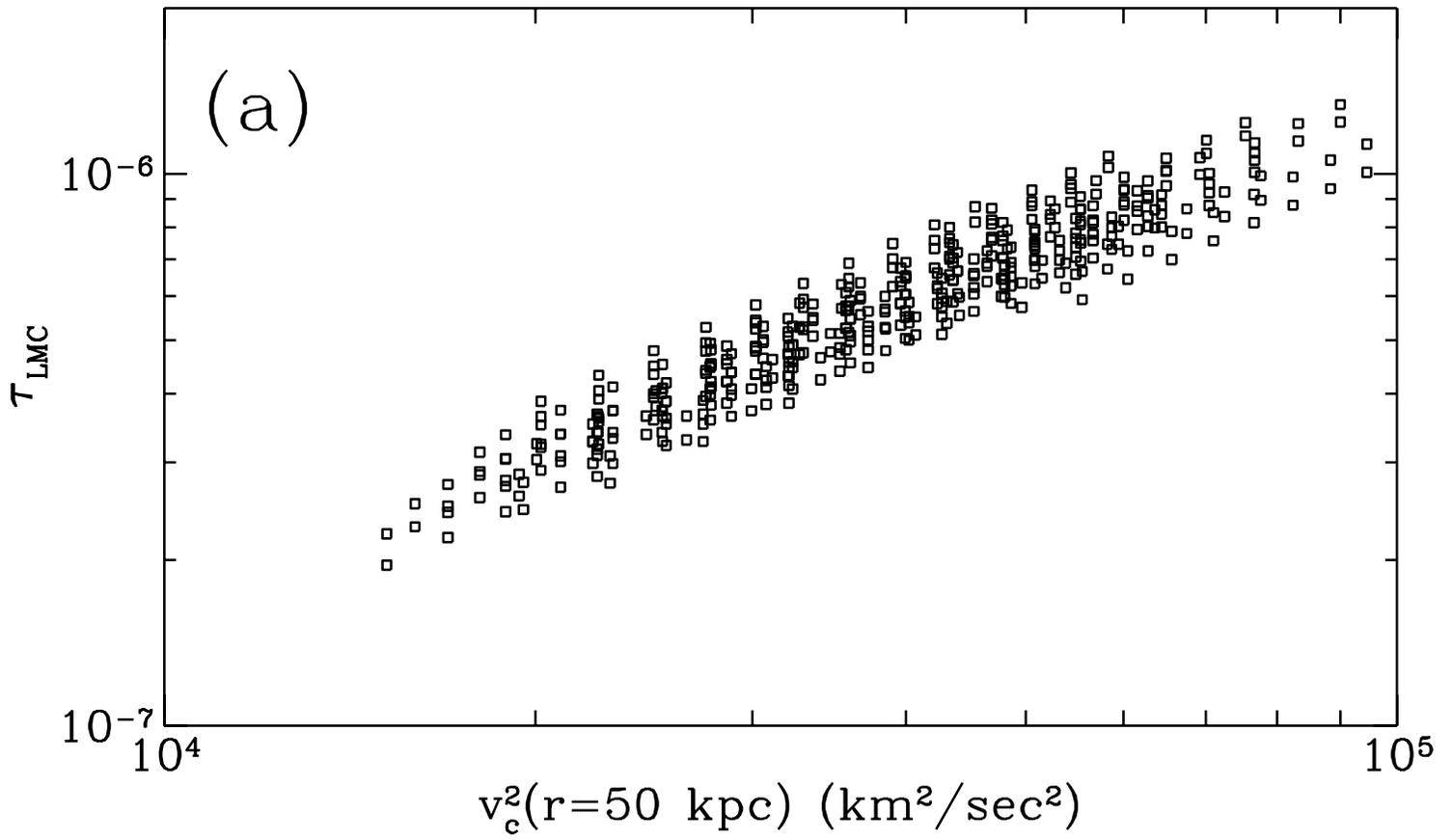
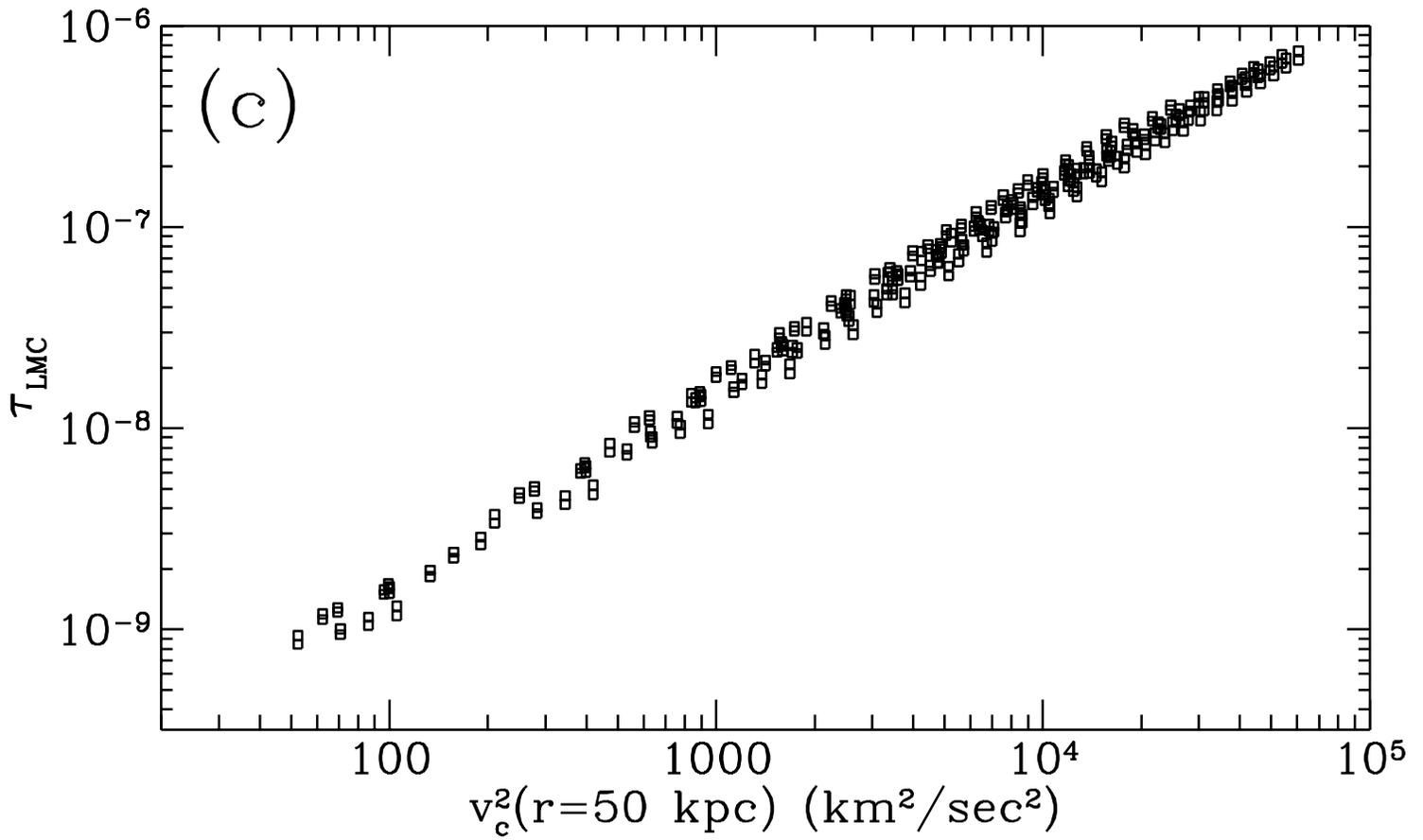
{\bf Figure 3:}
Scatter plots of optical depth vs $v_c^2$(50 kpc),
the square of the total rotation velocity at 50 kpc
in the galactic plane.  The mass of the Galaxy interior to this distance is
proportional to this squared velocity.  Parts (a) and (b) are for
a canonical disk, while parts (c) and (d) are for a maximal disk.
\medskip

\noindent
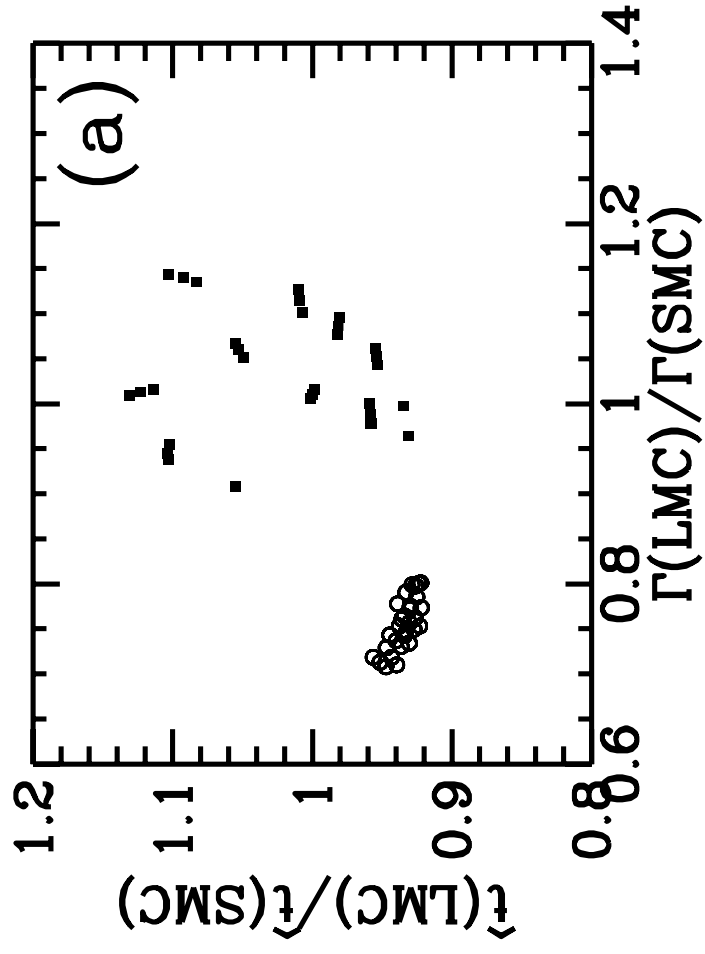
{\bf Figure 4:}
Finding the flattening parameter $q$.  Scatter plots of the ratio of
rates vs. the ratio of event durations.
The circles represent halo models which
are spherical ($q=1$), while the squares represent flattened halos
($q=.71$ for $\beta=0,-0.1$; $q=0.78$ for $\beta=0.1$).
part (a) is for LMC/SMC and shows clear separation of spherical and
flattened halos.
Part (b) is M31/LMC, and part (c) is M31/SMC.
\medskip

\noindent
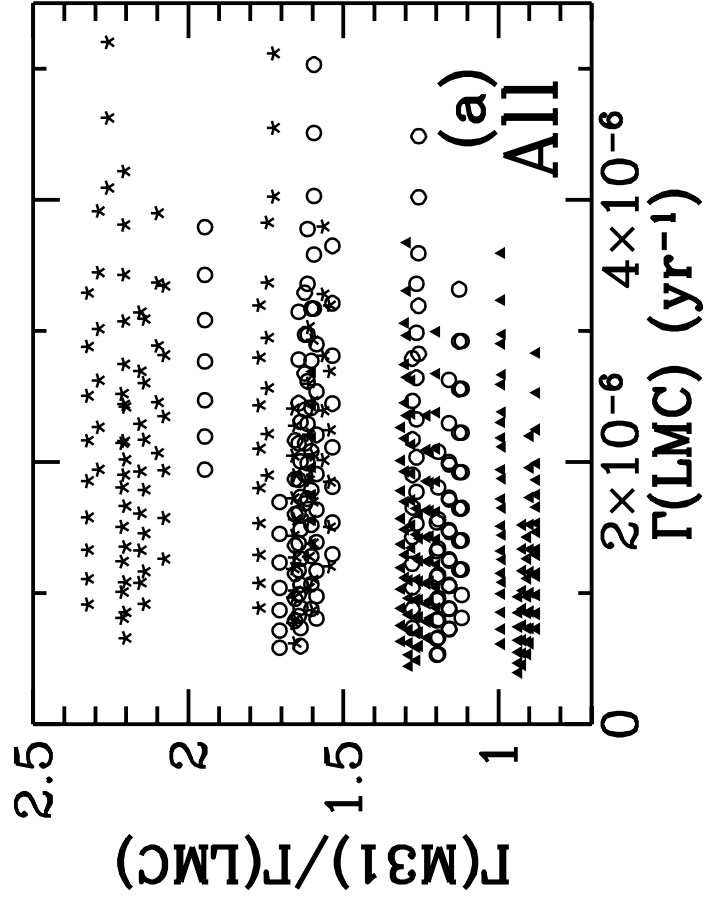
{\bf Figure 5:}
Finding the asymptotic slope $\beta$.
Scatter plots of the M31 rate divided by the LMC rate vs the LMC rate.
The stars represent halo models with $\beta=-0.2$ (rising rotation curve),
the circle models with $\beta=0$ (flat), and the triangles models
with $\beta=0.2$ (falling).
Part (a) shows all models, while part (b) shows only spherical models
and part (c) only the flattened models.  Separation of the models becomes
easier if the flattening is known.  The line of ambiguity in some
panels is due to $R_c=20$ kpc models, which can be distinguished
as shown in Figure 6.
All event rates
scale $\Gamma \propto \eta_m$.
\medskip

\noindent
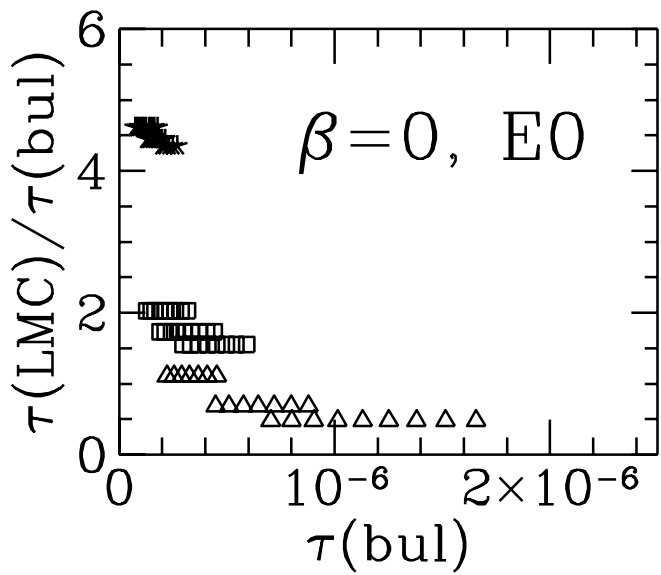
{\bf Figure 6:}
Finding the core radius $R_c$.
Each panel shows models of definite values of $\beta$ (rotation curve slope)
and $q$ (flattening).  The separation between models is quite good.
The triangles represent $R_c=2$ kpc, the squares represent $R_c=5$ kpc,
and the stars represent $R_c=10$ kpc.
Panels marked E0 are for spherical halos, while those marked E6 are
for flattened halos.
Models with $R_c=20$ kpc were
also considered but they are easily distinguished since they typically have
$\tau({\rm LMC})/\tau({\rm bul}) > 10$ and therefore fall off the top
of the figures.
\medskip

 \noindent
{\bf Figure 7:}
Examples of LMC $\that$ distributions for various model parameters.
The integral under each distribution is unity.
The curve marked (a) is for a ``standard" spherical halo
($\beta=0$, $q=1$, $R_c=5$ kpc, $v_0=200$ km/sec, $R_0=8.5$ kpc, and
$\Gamma=1.64 \times 10^{-6}$ events/yr).
Curve (b) has a shorter average duration
($\beta=-0.2$, $q=1$, $R_c=5$ kpc, $v_0=200$ km/sec, $R_0=8.5$ kpc, and
$\Gamma=3.9 \times 10^{-6}$ events/yr).
Curve (c) has a longer average duration
($\beta=0.2$, $q=0.78$, $R_c=10$ kpc, $v_0=210$ km/sec, $R_0=8.5$ kpc, and
$\Gamma=1.24 \times 10^{-6}$ events/yr).
Finally curve (d) has a maximal disk, which greatly reduces the amount of
halo material
($\beta=0$, $q=1$, $R_c=20$ kpc, $v_0=90$ km/sec, $R_0=7$ kpc, and
$\Gamma=9.37 \times 10^{-8}$ events/yr).
The average of each distribution is $\VEV{\that}$.
All event rates scale $\Gamma \propto \eta_m$.

\medskip
\vfill
\end